\documentclass{aa}

\usepackage{xcolor}
\usepackage{graphicx} 
\usepackage{appendix}
\usepackage{hyperref} 
\bibpunct{(}{)}{;}{a}{}{,} 

\title{Performance of AutoRegressive Planet Search methodology on simulated PLATO lightcurves}

\author{
Eric D. Feigelson\inst{\ref{PennState1}}   
              \inst{\ref{PennState2}}\inst{\ref{PennState3}}
\and Marco Montalto\inst{\ref{INAFCatania}}
\and Geert Jan Talens\inst{\ref{Oxford}}
\and Suzanne Aigrain\inst{\ref{Oxford}}
\and Luca Malavolta\inst{\ref{Padova1}}\inst{\ref{Padova2}}
\and Valerio~Nascimbeni\inst{\ref{Padova2}}
\and Tiziano Zingales\inst{\ref{Padova1}}\inst{\ref{Padova2}} 
}

\institute{
Department of Astronomy \& Astrophysics, 525 Davey Laboratory, Penn State University, University Park, PA 16802, USA \thanks{Corresponding author: \email{e5f@psu.edu}}  \label{PennState1}  
\and 
Center for Exoplanets and Habitable Worlds, 525 Davey Laboratory, Penn State University, University Park, PA 16802, USA \label{PennState2}
\and
Center for Astrostatistics, 525 Davey Laboratory, Penn State University, University Park, PA 16802, USA \label{PennState3}
\and
INAF, Osservatorio Astrofisico di Catania, Via S. Sofia 78, I-95123 Catania, Italy. \email{marco.montalto@inaf.it} \label{INAFCatania} 
\and 
Department of Physics, University of Oxford, Oxford, OX1 3RH, United Kingdom \email{geertjan.talens@physics.ox.ac.uk} \label{Oxford} 
\and 
Dipartimento di Fisica e Astronomia “Galileo Galilei,” Università di Padova, Vicolo dell’Osservatorio 3, 35122 Padova, Italy  \label{Padova1} 
\and 
INAF, Osservatorio Astronomico di Padova, Vicolo
dell’Osservatorio 5, 35122 Padova, Italy 
\label{Padova2}
}

\date{Received 17 December 2025 / Accepted 28 September 2026} 

\begin{document}

\abstract{} 
{This study examines the PLATO mission sensitivity to small transiting planets (radii between 0.5 and 2.0~R$_\oplus$) using Autoregressive Planet Search (ARPS) methodology based on simulated lightcurves.}
{Several thousand simulations of PLATO-like lightcurves from magnetically active stars expected from the PLATO Long Observation Phase South 2 (LOPS2) field with injected transiting planets smaller than 2.0~R$_\oplus$ radius are examined.  The lightcurves are analyzed with ARPS that uses autoregressive modeling for filtering of nonplanetary signals and the Transit Comb Filter periodogram for planetary transit detection.  False alarms of nonexistent periodic signals are reduced by thresholds on two ARPS parameters, peak periodogram power, and transit duration.}
{When applied to PLATO's simulated first three months observations of bright stars (P1 and P2 samples), at least 50\% of Earth-size planets are recovered for periods shorter than 40~d. In two years of LOPS2 observation, 25\% are recovered for periods less than 240~d.  Methodological implications on detrending, periodograms, False Alarms, vetting, and catalog evaluation are discussed.  The utility of a positive predicted value heat map after detection and vetting is demonstrated.}
{While the ARPS planet detection procedure is not entirely well-suited for PLATO analysis, this study provides methodological insight into the sensitivity and reliability of small transiting planet detection.}

\keywords{methods: statistical -- planets and satellites: detection -- surveys -- techniques: photometric}

\titlerunning{ARPS Performance on Simulated PLATO lightcurves}
\authorrunning{E. D. Feigelson, M. Montalto, et al.}

\maketitle
\nolinenumbers

\section{Introduction} \label{intro.sec}

The discovery of Jupiter-size planets orbiting close to solar-type stars, first by radial velocity variations and then by transit photometry, inaugurated exoplanetary research \citep{Mayor95, Charbonneau00}.  Hot Jupiters are easily found today and  nearly 10\% of the $\simeq 6000$ confirmed planets\footnote{\textcolor{red}{NASA Exoplanet Archive, September 2025.}} have radii above 1~R$_{\rm Jup}$ and orbital periods shorter than 10~days despite a much lower unbiased occurrence rate around 1\% \citep{Howard12}.  

However it is much more challenging to detect Earth-size planets in Earth-like orbits around solar-type stars; only a handful of confirmed planets have radii consistent with rocky compositions and stellar insolations consistent with liquid water on their surfaces.  The {\it Kepler}\footnote{\url{https://science.nasa.gov/mission/kepler/in-depth/}} and {\it TESS} \citep{Ricker14} missions found it challenging to identify significant samples of habitable planets despite their stated goals. 

A major cause of these difficulties has been magnetic activity signatures that inhibit high precision measurements in both photometric and radial velocity.  For transit photometry with the {\it Kepler} satellite observatory, intrinsic stellar variability for an ensemble of stars roughly doubled the instrumental noise \citep{Gilliland11}.  For the {\it Kepler} and {\it TESS} (Transiting Exoplanet Survey Satellite) missions \citep{Borucki10, Ricker14}, elaborate analysis procedures were designed to reduce stellar and instrumental variations and to minimize false alarm (when a periodic signal is incorrectly reported) and false positive (when a non-planetary periodic signal is recovered) signals  \citep{Jenkins17, Guerrero21}.  The goal is to improve the sensitivity and reliability of official catalogs of candidate transiting exoplanets emerging from dedicated space-based surveys. 

The European Space Agency's PLAnetary Transits and Oscillations of stars (PLATO) space telescope planned for an early-2027 launch is designed to reduce these limitations to `detect small planets ($<2$~R$_\oplus$) around bright stars ($P_{\rm mag} <$11 mag), including terrestrial planets in the habitable zone of solar-like stars.'  \citep{Rauer25}.  It will obtain continuous lightcurves for more than 245,000 target stars lying in a single Long-duration Observations Phase South 2 (LOPS2) field for at least two years  \citep{Montalto2021, Nascimbeni25, Montalto26}.  

Sensitivity to Earth-like transits requires highly effective statistical techniques to model and remove intrinsic stellar photometric variations without also diminishing the faint and widely-spaced periodic dips from long period, potentially habitable, transiting planets. Filtering techniques must be nonparametric without a global statistical model because stellar variability is largely unpredictable with components on many timescales.  Statistical methods for filtering include sliding windows (including differentiation), polynomial fitters (such as splines and locally estimated scatterplot smoothing), Gaussian Processes regression with carefully chosen kernels, multiscale wavelet transforms, Fourier filtering, self-organizing maps, independent components analysis, recurrent and other neural networks, low-dimensional autoregressive modeling (ARMA), and other procedures. Comparisons of some filtering methods are made by \citet{Hippke19}, \citet{Canocchi23}, and \citet{ Garcia24}. 

The periodic transit brightness dips are then sought in the residuals of the detrended lightcurves, most commonly using the Box-Least Squares (BLS) algorithm or its variants \citep{Kovacs02, Collier06, Ofir08, Ofir14, Mislis16, Hippke19}. However, if filtering involves differentiation of the lightcurve, a transit box shape is transformed to a double-spike and the Transit Comb Filter (TCF) periodogram is used \citep{Caceres19a}.  Tests show that TCF outperforms BLS under many circumstances \citep{Gondhalekar23}. 

\citet{Heller2022} and \citet{Krenn24} quantitatively examine the sensitivity of PLATO to Earth-sized planets.  They introduce both planetary transits and stellar noise based on short-term solar variability into simulated lightcurves based on PLATO instrumental capabilities.  Using filtering with the {\tt W\~otan} procedures \citep{Hippke19} and post-filtering transit detection with the transit least-squares algorithm, \citet{Heller2022} find true positive rates near unity for planets larger than $1.2 [1.0]$~R$_\oplus$ with two[three] or more transits in 2--3~yr continuous observation. \citet{Krenn24} examine four statistical procedures for quantifying correlated noise combined with a BLS-type algorithm that includes limb-darkening in the host star.  They find that, for targets brighter than $P_{\rm mag}$ of 10, a transiting Earth analogue can be reliably identified if more than 2 transits are observed; its radius can be measured with about 3\% precision. A more extensive comparison of methods for PLATO transit detection is provided by \citet{Talens26}.  

The purpose of the present study is to provide complementary calculations of filtering and periodogram procedures for the recovery of transits injected into noisy lightcurves.  Here we use statistical techniques adopted from the AutoRegressive Planet Search (ARPS) procedure developed by \citet{Caceres19a} with extensions by \citet{Melton24a,  Melton24b}.  These statistical procedures have been applied to {\it Kepler} \citep{Caceres19b} and {\it TESS} \citep{Melton24c} survey lightcurves.  We apply these methods to simulations of PLATO lightcurve simulations with different assumptions of stellar photometric variability (Sect. \ref{simulations.sec}).  

The goals of the present study are limited as follows:
\begin{enumerate}
    \item Only ARPS, one of several existing lightcurve analysis systems, is considered for transiting planet detection.  It is mathematically different from the planet detection procedures implemented in the PLATO pipeline \citep{Rauer25}. 

    \item Only recovery of a single small planet (radius between 0.5  and 2.0~R$_\oplus$) is considered.  Large planets and multiple planets systems are not treated. 

    \item The methodology is not designed for detection of single transits \citep[e.g.][]{Cooke18}.  
    
    \item We consider sensitivity to transiting exoplanets in the presence of photon noise, instrumental gaps in the data stream, and modeled stellar variability.  Statistical removal of blended eclipsing binary contamination is not tested. These are discussed by \citet{Montalto20} and \citet{Melton24b} for the {\it TESS} mission and by \citet{Bray23} and \citet{Smith25} for the PLATO LOP2S field.  In statistical parlance, we treat false alarms (when no periodic signal is truly present) but not false positives (periodicity not caused by planetary transits is present). 
\end{enumerate}

Sect. \ref{simulations.sec} describes the simulations used in this study. In Sect. \ref{arps.sec} and Appendix A, we review the ARPS methodology. Section~\ref{1Q.sec} presents the expected sensitivity of the PLATO mission from its first quarter of observations; this is extended to 2 years near-continuous observation of the LOPS2 field in Sect. \ref{2yr.sec} and Appendix B. In  Sect. \ref{discussion.sec}, we discuss the results of the analysis, relationship to PLATO pipeline methods, and prospects for improvements. Appendix C discusses the role of multivariate feature-based classifiers that can improve the recovery rates obtained here.

\section{Simulated datasets} \label{simulations.sec}

The simulated lightcurves were generated by the PLATO consortium for the purpose of comparing different lightcurve filter and transit search algorithms, expanding the work presented in \citet{Canocchi23} to planets with smaller radii and longer periods. The simulated lightcurves are described in detail by \citet{Talens-2026a} and are available from the Oxford research archive\footnote{\url{https://ora.ox.ac.uk/objects/uuid:682bcbb6-ed68-42ca-8c00-277a3aea64bd}}. We briefly summarize the key features here.

The stellar parameters for the simulations were taken from the P1 and P2 samples (13330 stars) PLATO input catalog (PIC) for the LOPS2 field \citep{Montalto2021, Nascimbeni25, Montalto26}. P1 consists of F5-K7 dwarfs and subgiants with $V$ brighter than 11 and noise below 50~ppm, while P2 is further restricted to $V$ brighter than 8.5. Stars in the PLATO P1 and P2 samples  were selected and cuts were applied to: surface gravity ($\log g$ above  4) to exclude giant stars; stellar effective temperature (T$_{\rm{eff}}$ between 4594 and 6334~K) due to limitations in the code used to generate the spot signals; and stellar magnitude ($V$ brighter than 8) due to limitations in the code used to generate noise. After applying these cuts, about half the sample (6552 stars) remains from which we randomly selected 2000 stars to be simulated. Stars were randomly assigned an inclination ($i_\star$) and, with an assumption of solar metallicity for all stars, four-parameter limb-darkening coefficients are obtained from \citet{Kostogryz-2022}. 

\begin{figure}
\resizebox{\hsize}{!}{\includegraphics{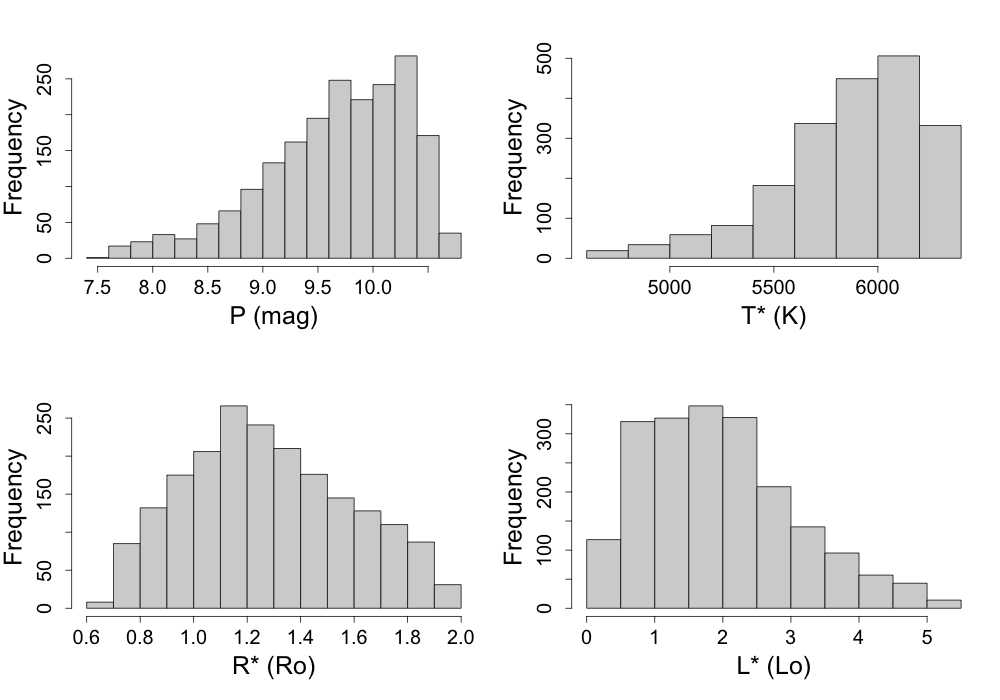}}
\caption{Histograms of input parameters from the PIC of the LOPS2 field for 2000 stars used in the PLATO simulations outlined in Sect. \ref{simulations.sec}. Distributions represent a subsample of the P1 + P2 stars from the PLATO Input Catalog for LOPS2 \citep{Montalto2021, Nascimbeni25}. \label{OxPad_input_star_hist.fig} }
\end{figure}

The simulations begin by generating the effects of granulation, oscillations, random and systematics noise for the 2000 selected stars using version 1.8 of the PLATO solar-like lightcurve simulator \cite[PSLS;][]{Samadi-2019}. Simulations were generated at PLATO's native 25-second observing cadence, and assume a nominal 2-year pointing composed of quarter lengths ranging from 84 to 97 days. Each quarter starts with a 2-day gap in observations for spacecraft roll and commissioning observations, and 20 minute gaps are added every $72\pm2$ hours representing spacecraft momentum dumps. Out-of-transit variability due to star spots were then added to these PSLS lightcurves using a new version of {\tt pyspot}\footnote{\url{https://github.com/talensgj/pyspot}} \citep{Aigrain-2015}. It uses the spot model from \citet{Kipping-2012} that accounts for limb-darkening.

Two versions of each lightcurve were created, one containing a planet on an orbit interior to the habitable zone (inner simulations) and one containing a planet on an orbit inside the habitable zone (outer simulations). This produced two sets of 2000 lightcurves where only the planet signals differ. Definitions of the habitable zone were taken from \citet{Kopparapu-2013} with their Recent Venus and Early Mars scenarios setting the inner and outer edges of the habitable zone, respectively. The luminosity of a main sequence star with the same effective temperature was used rather than the actual luminosity from the PIC when calculating the habitable zone boundaries; otherwise PIC luminosities would result in inner boundaries close to the 2-year pointing duration that would not be practical for these tests. Shrinking the habitable zone means that the inner and outer simulation labels are not realistic, but this should have little effect on the comparison of different algorithms in recovery tests of Earth-size planets. 

The injected signals correspond to planet sizes between 0.5 and 2.0 R$_\oplus$, impact parameters between 0 and 1, and circular orbits. For the inner sample, each planet was assigned an orbital period between 1.5 times the Roche limit of the star and the inner boundary of the habitable zone. For the outer sample, orbital periods lie between the inner and outer boundary of the habitable zone up to a maximum period of 700 days. The epoch of first transit was randomly assigned  between 0 and the orbital period for the inner simulations.  For the outer simulations, the first transit epoch is random out to $P \sim 368$~days (guaranteeing two transits in the 2-year lightcurve) and, for longer periods, was forced to lie within the first quarter to ensure at least 1 transit is observed by PLATO. Transit durations range from 0.02~d to 1.0~d;   \citet{Talens25} gives the dependency of transit durations on orbital period.  

Fig.~\ref{OxPad_input_star_hist.fig} shows the distribution of input parameters for the 2000 stellar hosts. These reflect a subset of the P1 + P2 sample from the PIC for the LOPS2 field \citep{Montalto2021,  Montalto26, Nascimbeni25}.

\section{Autoregressive planet search methodology} 
\label{arps.sec}

The ARPS procedure for PLATO simulations begins by binning the lightcurves from the native 25-second cadence to a continuous 30 minute cadence. This arranges for ingress and egress changes to be sudden rather than gradual; this improves periodogram sensitivity.   Next ARFIMA (AutoRegressive Fractional Integrated Moving Average) modeling is applied to the binned lightcurves, a combination of a differentiation operation for filtering followed by a low-dimensional nonlinear parametric modeling of stochastic autocorrelated time series.  

Commonly known as Box-Jenkins analysis, ARMA-type modeling  was developed around 1970 for analysis of econometric time series such as variations in the value of stocks.  These autoregressive models have been used in a vast range of time series and signal processing problems; a nonlinear extension for volatility received the Nobel Prize in Economic Science \citep{Engle82}.  The general use of autoregressive modeling for astronomical time series is discussed by \citet{Feigelson18} and the specific use for filtering stellar lightcurves for transiting planet detection is presented by \citet{Caceres19a} and \citet{Melton24a}.  We recommend the online textbook by \citet{Hyndman21} for explanations of autoregressive modeling including the ARFIMA model. 

The first step (the `I' or 'Integrated' in ARFIMA) is the differencing operator, the differentiation of the evenly spaced time series where the stellar flux value at time $x_t$ is replaced by $x_t - x_{t-1}$.  This is equivalent to removing a narrow running mean filter.  For a regularly spaced time series $x_t$, let $d_t = x_t - x_{t-1}$ be the differenced time series and $m_t=\frac12(x_t + x_{t-1})$ be the running mean with window of 2 time units. Then $m_t = \frac12 d_t + x_{t-1}$. One can thus recover the running mean local smoother with window 2 from the differenced local change operator by dividing by 2 and adding the lag-1 offset series.  The running mean is the simplest local polynomial smoother; a spline regression typically uses a higher order cubic polynomial. All polynomial smoothers will not fully model structure smaller than the window width.  

For most real {\it Kepler} and {\it TESS} stellar lightcurves, and nearly all simulated PLATO lightcurves, the differencing operator removes the nonstationary flux variations due to magnetic activity, stellar rotation, and other secular or quasi-periodic variation.  

The second step is to reduce short-memory stochastic autoregressive behaviors that are not removed by differentiation.  The simplest linear model for a stationary, evenly-spaced time series is an autoregressive moving average (the `ARMA' in ARFIMA) model. Here the current flux value is modeled as the sum of linear dependence on $p$ recent previous values of the flux and $q$ recent previously values of stochastic flux changes, ARMA($p,q$).  Letting $x_t$ be the current value and $\epsilon_t = N(0, \sigma^2)$ be a normally distributed error associated with time $t$, then the ARMA($p,q$) model is 
\begin{eqnarray}
x_t = \theta_1 x_{t-1} + \theta_2 x_{t-2} + ... + \theta_p x_{t-p} + \epsilon_t + \\ \nonumber 
\phi_1 \epsilon_{t-1} + \phi_2 \epsilon_{t-2} + ... + \phi_q \epsilon_{t-q}.
\end{eqnarray}

The ARMA parameters  $\theta_{i=1,...,p}$, $\phi_{j=1,...,q}$, and $\sigma^2$ are fit by maximum likelihood estimation for a range of models with a range of integer values of $p$ and $q$ representing models of different order.  The optimal values of $p$ and $q$ maximizes a likelihood ratio statistic penalized for model complexity.  Typically the Akaike Information Criterion (AIC) is used for ARMA model selection  \citep{Konishi08, Box15, Hyndman21}.  The adequacy of the AIC-selected model is established with the theorem-based Ljung-Box test \citep{Ljung78} of the residuals.  A p-value greater than 0.01 residuals for this tests demonstrates the residuals are consistent with stationary (trend-free) Gaussian white noise and no further modeling is needed. 

The fitting procedure is thus deterministic with no free parameters such as a choice of kernel function or width.  The best fit ARMA(p,q) model has the effect of removing most short-memory stochastic autocorrelated ripples in flux that remain after the differencing operation removed large-scale variations.  

As the linear ARIMA (AutoRegressive Integrated Moving Average) model sometimes fails to remove  autocorrelations on longer timescales, we apply a nonlinear ARFIMA model (the `FI' represents  `fractional integration') that is equivalent to adding a long-memory power law $1/(f^\alpha)$ component to the short-memory autoregressive model. The mathematics relating fractional integration to power law `red noise' is presented by \citet{Palma07}.  We find that only $\sim$10\% of fits to PLATO simulations require a long-memory component with $\alpha$ greater than 0.02.  

A guide to the ARPS software is provided in Appendix~A.   Figure~\ref{app_ARFIMA.fig} in Appendix~B shows three examples of ARFIMA modeling of simulated PLATO lightcurves. In two cases, the ARFIMA model was completely successful in removing all simulated stellar variations.  In the third case, residual autocorrelation remains in the ARFIMA residuals implying that a more complex model might further reduce autocorrelated noise.  

\begin{figure*}
\centering
\includegraphics[width=17cm]{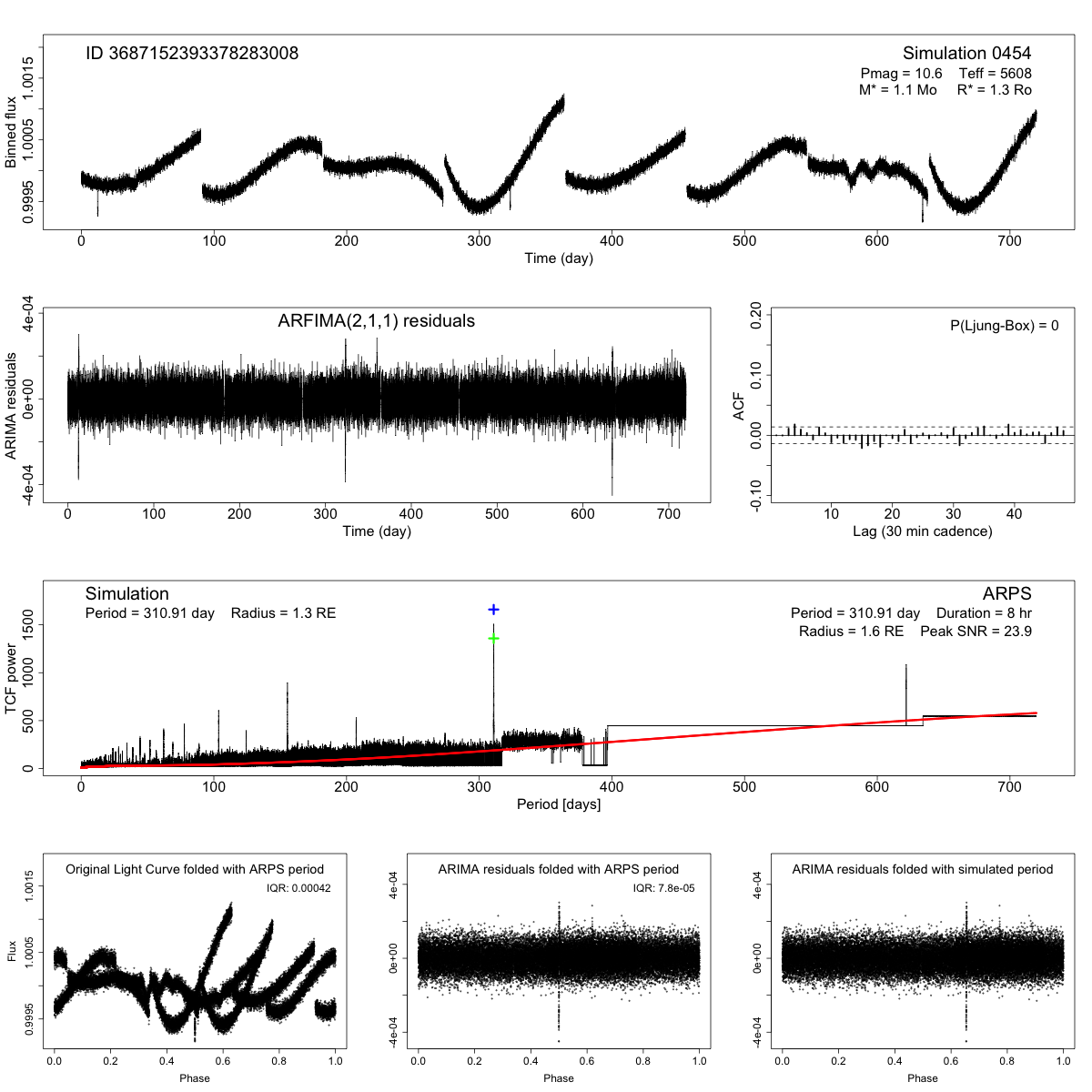}
\caption{ARPS graphical output for a recovered simulated transit planet signal in the outer simulations. ACF is the autocorrelation function as a function of lag in units of the 30~min cadence.  In the TCF periodogram, the green $+$ symbol represents the simulated injected period while the blue $+$ symbol is the recovered ARPS period. S/N is the signal to noise ratio.} \label{ARPS_example_good.fig}
\end{figure*}

\citet{Caceres19b} and \citet{Melton24a} find that the residuals of ARIMA and ARFIMA models for most (but not all) {\it Kepler} and {\it TESS} lightcurves are close to Gaussian white noise.  A critical finding is that ARIMA-type modeling is not effective for removing sudden, single-bin spikes like stellar flares or transiting planet ingresses and egresses.  The differencing operator changes a periodic box-shaped transit signal into a periodic sequence of down-and-up spikes representing the sudden ingress and egress flux changes.  These signals are prominent only if the lightcurve cadence is similar to the duration of the ingress, roughly 20--30~min for most exoplanets (Sect. \ref{limitations.sec}).  

The ARFIMA residuals are thus typically stationary and near-Gaussian white noise, but still containing sudden signals like flares and transits. Planet detection then requires the sensitive detection of sequences of double spikes in ARIMA residuals. \citet{Caceres19a} developed the TCF for this purpose.  Once an orbital period, transit duration, and phase are chosen, the algorithm passes a matched filter for periodic double-spikes over the lightcurve ARIMA residuals. A periodogram is then constructed with the maximum power at each period; this is similar to the BLS algorithm but without a least-squares calculation.  \citet{Gondhalekar23} compared TCF and BLS sensitivities for a variety of simulated and observed lightcurves.  They show that BLS is often subject to higher trends, noise and spurious spectral peaks than TCF.  

\begin{figure*}
\centering
\includegraphics[width=17cm]{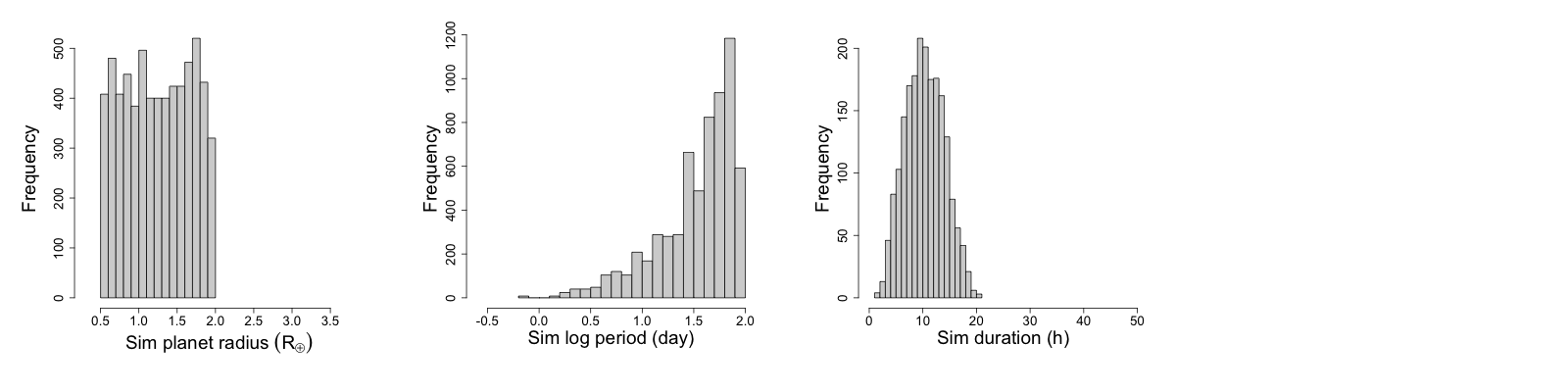}
\includegraphics[width=17cm]{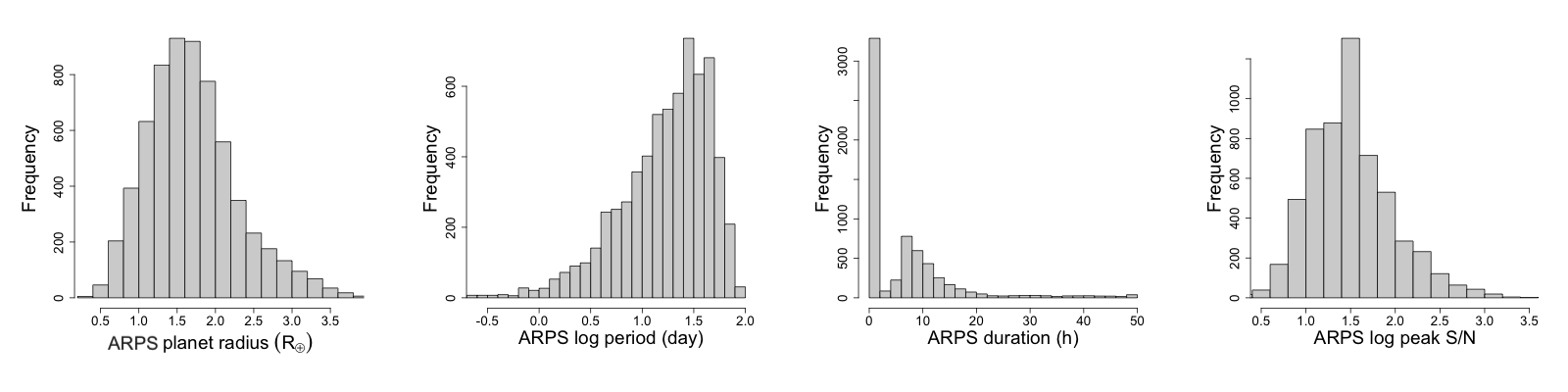}
\caption{Comparison of input simulation parameters with derived ARPS parameters for one-quarter simulations. Input parameters are in the first row and derived parameters are in the second row.   \label{OxPad_1Q_sim_hist.fig} }
\end{figure*}  

\begin{figure*}[ht]
\centering
\includegraphics[width=17cm]{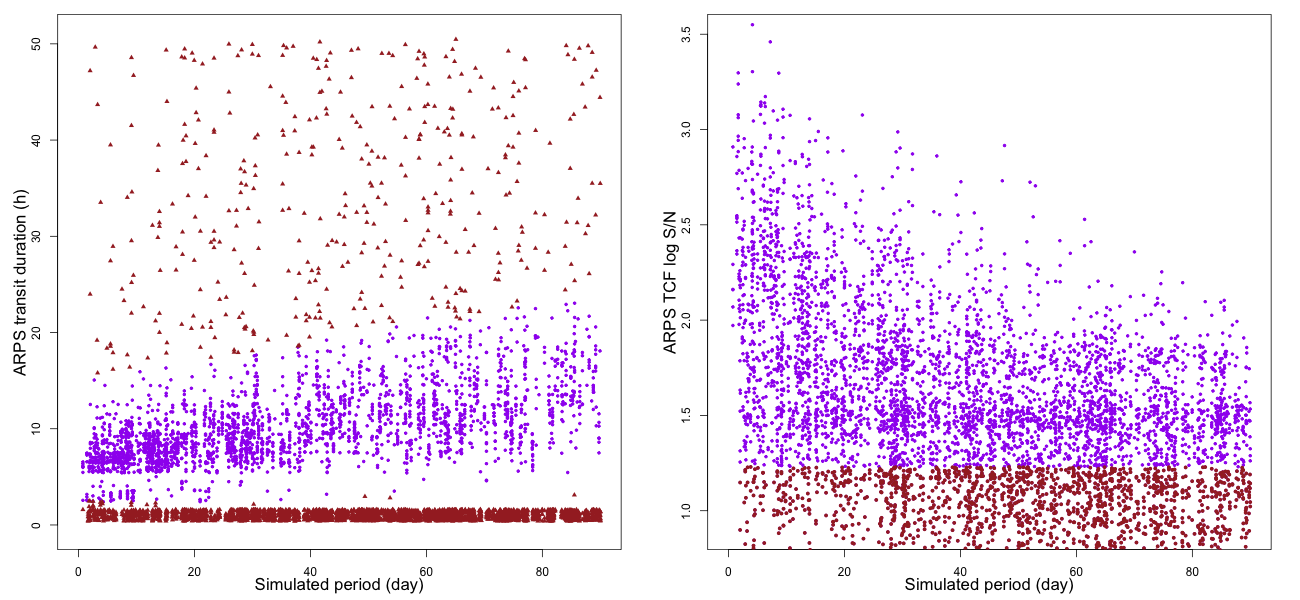}
\caption{Heuristic cleaning of ARPS TCF periodogram best peaks in one-quarter segments of the inner orbit simulations. Left panel shows removals due to unlikely transit durations, and right panel shows removals due to weak periodogram peak S/Ns. Blue  points are accepted peaks and green points are rejected.  \label{OxPad_1Q_cleaning.fig}}
\end{figure*}

The software implementation of the ARPS transit detection procedure $-$ ARFIMA modeling for filtering and TCF periodograms $-$ is computationally efficient and simple to code.  Appendix A provides software details and guidance for researchers.  Relevant code\footnote{\url{https://github.com/Yash-10/Periodogram-Comparison-Optimize-Planet-Detection}} developed by \citet{Gondhalekar23} is available at \citet{Gondhalekar23b}.  Computational time on an Apple M4 CPU for a simulated 2-year PLATO lightcurve, binned to 30-min cadence, is about 20 seconds for ARFIMA, 2 minutes for the TCF periodogram (with periods starting at 0.2~day), and less than 1 minute for other calculations. TCF computation scales as $O(D^3)$ where $D$ is the duration of the lightcurve.

Fig.~\ref{ARPS_example_good.fig} shows an example of successful recovery of a 1.3~R$_\oplus$ radius planet in the outer simulations.  The top row shows the simulated 2-year lightcurve, binned to a 30-min cadence.  It is dominated by assumed approximately 0.1\% amplitude photometric variations and jumps between quarters when satellite pointing is reconfigured.  Transits around times 10~d, 320~d and 630~d are visibly present in the simulated lightcurve.  

The second row shows the residuals from an ARFIMA(2,1,1) model chosen with maximum likelihood estimation and AIC model selection.
The notation ARIMA({\it p,d,q}) or ARFIMA({\it p,d,q}) denotes the order, or complexity, of the autoregressive model.  $p$ is the number of coefficients in the autoregressive component, $q$ the number of coefficients in the moving average component, and $d$ is the number of filtering differencing operations performed.  $d$ is real-valued for fractional differencing in ARFIMA models.  We enforce a single differencing operation so a transit box-shaped dip in the lightcurve becomes a double-spike that can be found with the TCF periodogram. But $p$ and $q$ values are selected using the AIC penalized likelihood criterion.  The residuals are nearly consistent with Gaussian white noise, although the Ljung-Box test shows statistically significant autocorrelation is still present. The three transit signals are now seen as double-spikes.  

The second row also shows the autocorrelation function (ACF) of the model residuals.  Here the dash lines represent the 95\% confidence band for Gaussian white noise.   The autocorrelation function of the residuals shows little structure although the sensitive Ljung-Box test indicates it is inconsistent with Gaussian white noise.

The third row of Fig.~\ref{ARPS_example_good.fig} shows the TCF periodogram calculated for periods ranging from 0.2 to 720 days. The simulated transit period of 310.91 day  is accurately recovered by the strongest peak in the TCF periodogram.  Expected alias peaks at factors of 2, 1/2, 1/3, ... of the true period are also present.  The bottom row shows the original and ARIMA residuals folded with the peak TCF power period and simulated period.  The expected signal for a true transit is a narrow double-spike in the middle panel. TCF periodograms are well-behaved for most simulations with transit signals at shorter periods. Mild trends (shown with a LOESS fit in red in the third row of Fig.~\ref{ARPS_example_good.fig}) and heteroscedasticity in periodogram noise values are seen; these behaviors are discussed by \citet{Ofir14} and \citet{Gondhalekar23}.

Deterioration of periodogram information is seen at periods longer than half the duration of the lightcurve.  Noise values often cannot be reliably measured so peak signal-to-noise (S/N) values may not be meaningful. Spurious peaks associated with quarter gaps or residual autocorrelations can also appear.  Care is needed in interpretation of long-period peaks in TCF and other periodograms. Other examples of periodogram deterioration are shown in Appendix~\ref{AppendixB.ARPSperformance}.

\section{Analysis and results}

\subsection{Single quarter PLATO analysis} 
\label{1Q.sec}

For the estimation of PLATO sensitivity of data from one-quarter (90~d) simulated lightcurves, each inner orbit simulation is divided into eight single-quarter segments that are subject to independent ARPS analysis.  Figure~\ref{OxPad_1Q_sim_hist.fig} compares the distributions of three simulated planetary parameters $-$ planet radius, orbital period, and transit duration $-$ with the results of the ARPS analysis for the one-quaarter simulations.  Under certain conditions, ARPS can overestimate planetary radii, underestimate planetary orbital periods, and derive transit durations that are too short or too long \citep{Caceres19b, Melton24b}.  

While the differences between simulated and recovered transiting planet properties seen in Fig.~\ref{OxPad_1Q_sim_hist.fig} might initially suggest that recovery of small planets will fail, the comparison of histograms suggest opportunities for discriminating between spurious false alarms and correct discovery of true positives.  For example, the rejection of TCF peaks with transit durations shorter than 1 hour and longer than $\simeq$20 hours should clean away many false alarms.  Furthermore, experience with TCF performance with {\it Kepler} and {\it TESS} data, as well as {\it TESS}-based simulations, suggests that TCF peaks with S/N less than approximately 20 are often spurious \citep{Caceres19b, Gondhalekar23, Melton24a}.  

Fig.~\ref{OxPad_1Q_cleaning.fig} shows our chosen thresholds for two derived parameters: the transit duration associated with the TCF peak and the S/N of the TCF peak. These thresholds seem reasonable, but the cuts are not based on any algorithmic or statistical procedure. Other variables could be used for cleaning the sample, such as TCF transit depth.  More effective removal of false alarms obtained with machine learning algorithms trained on many variables is discussed in Appendix C.

\begin{figure*}[ht]
\centering
\includegraphics[width=17cm]{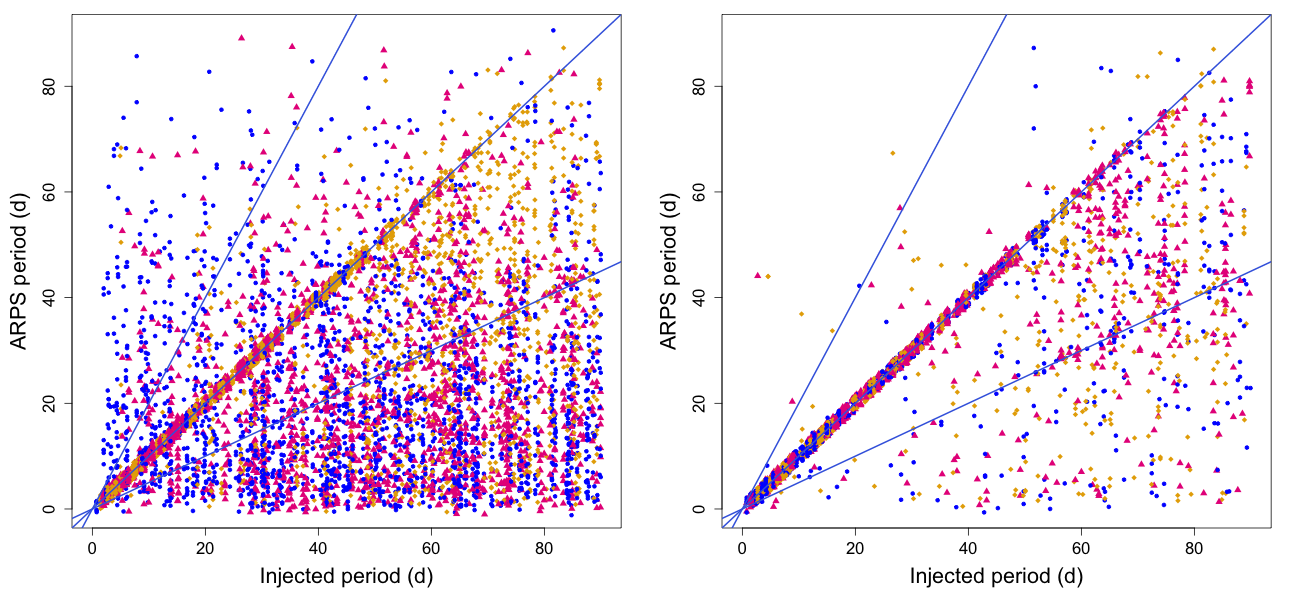}
\caption{Recovered periods $vs.$ simulated periods (in days) for ARPS analysis of single quarter PLATO simulations.  Left: Full samples from the ARPS processing.  Right: After removal of likely false alarms (red points in Fig.~\ref{OxPad_1Q_cleaning.fig}).  Symbols are coded by the injected planet radius in Earth radii: blue circles for 0.5--1.0~R$_\oplus$; red triangles for 1.0--1.5~R$_\oplus$; and gold  diamonds for 1.5--2.0~R$_\oplus$.   The central blue line represents correct identification of the simulated period by ARPS; the outer lines represent 2-times and 1/2-times aliases.  Points are jittered to reduce overlapping symbols.  \label{1Q-periods.fig} }
\end{figure*}

\begin{figure*}[ht]
\centering
\includegraphics[width=8.8cm]{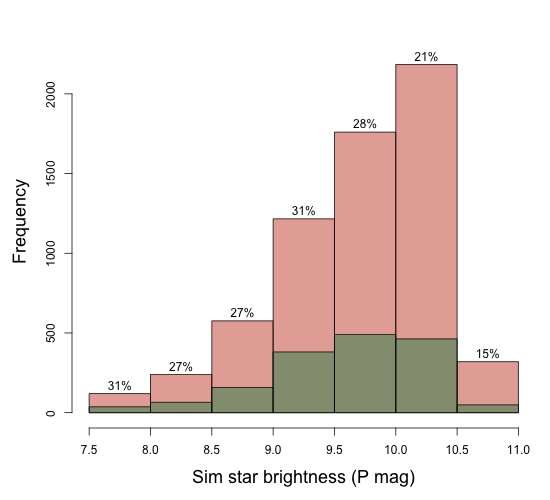}
\includegraphics[width=8.8cm]{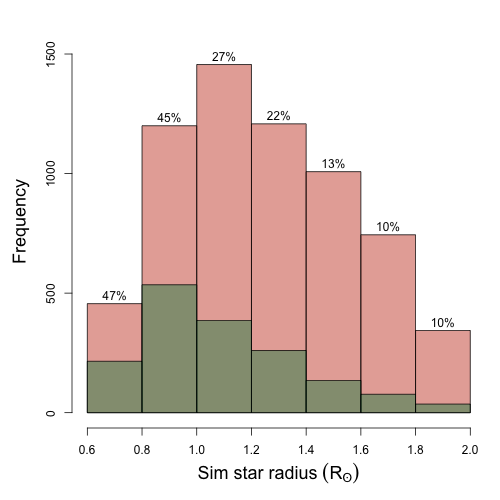} \\
\includegraphics[width=8.8cm]{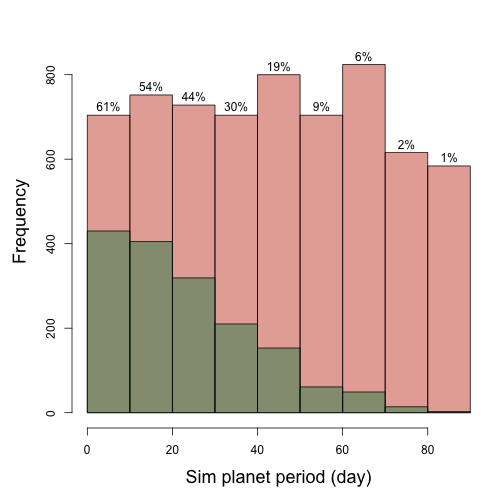}
\includegraphics[width=8.8cm]{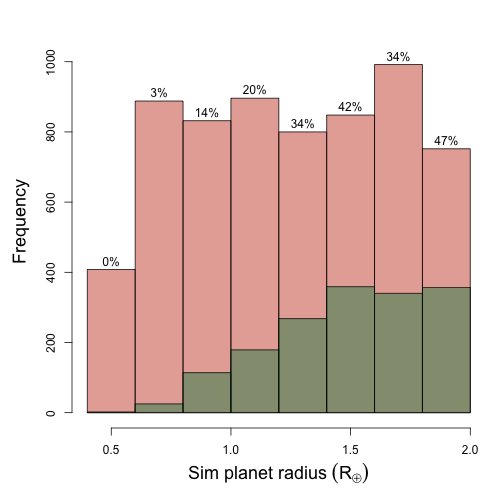}
\caption{ARPS recovery for injected planets for four input parameters of simulated single-quarter PLATO lightcurves.  The green histograms represent lightcurves where ARPS successfully recovered the injected planetary period after false alarm removals (Figure~\ref{1Q-periods.fig}-right).  The green histograms represent the full sample.  From upper left to bottom right, the properties are: star brightness in $P$ magnitude, and star radius in R$_\odot$, simulated orbital period in days, simulated planet radius in R$_\oplus$.  } \label{Four_1Q_hist.fig}
\end{figure*}

\begin{figure*}
\centering
\includegraphics[width=17cm]{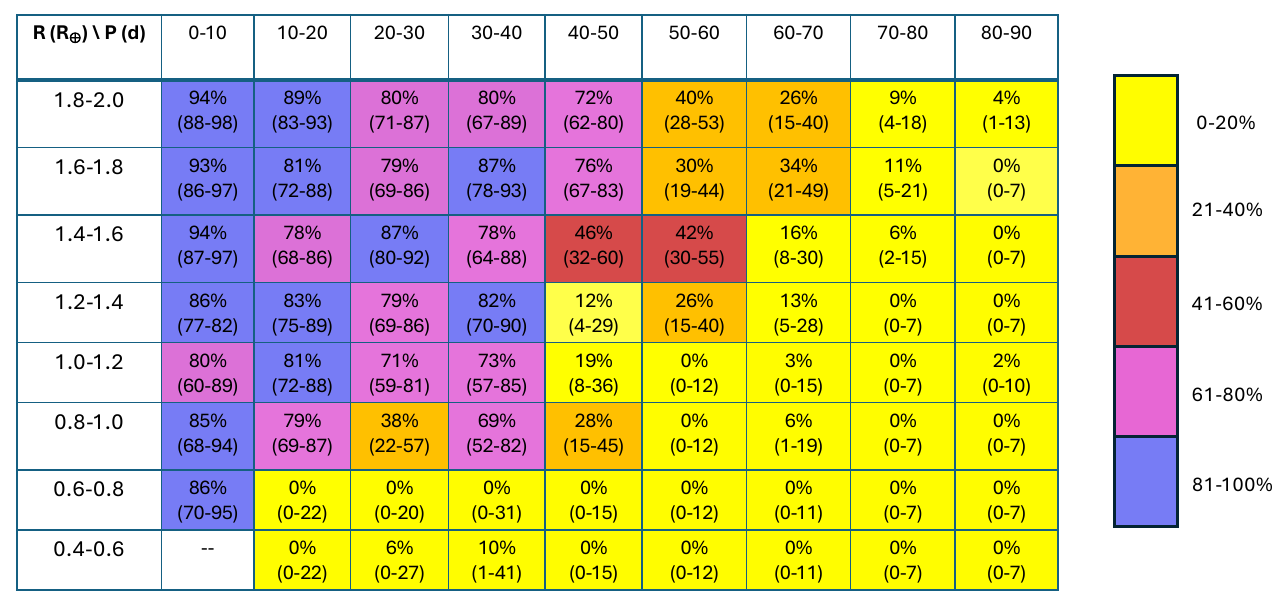}
\caption{Heat map for one-quarter PLATO planet recovery using ARPS methodology as a function of orbital period in days (columns) and planet radius in R$_\oplus$ (rows).  The value in each radius-period cell gives the percentage of simulated planets recovered based on the heuristic cleaning shown in Fig.~\ref{1Q-periods.fig}. The 90\% confidence interval of percentages, based on the binomial distribution and dependent on the number of simulations in each bin, is given in parentheses.  Bin colors reflect the gradient in sensitivity as indicated by the color bar at right. An uncolored cell with dash indicates too few simulations are available so the confidence interval exceeds 50\%. } \label{1Q_recovery_table.fig}
\end{figure*}

The result of this cleaning process can be seen by comparing the left and right panels of Fig.~\ref{1Q-periods.fig}: a much higher fraction of ARPS periods accurately agree with the input simulated periods once transit duration and S/N cuts are implemented.  For periods shorter than approximately 45 ~days, about 95\% of the TCF spectral peaks are correctly with TCF periods accurate to $\pm$0.06\% (1$\sigma$). From  45  to 90~days, decreasing fractions of the TCF peaks are correctly identified.  This is partly due to the increasing presence of only one transit (hence no signal in the TCF periodogram) and partly to the paucity of photometric measurements during widely-spaced transits.  In cases where ARPS derives incorrect periods, the TCF period is shorter than the true injected period.  In a few cases, ARPS captures the 1/2 alias of the simulated period.  

\begin{figure*}
\centering
\includegraphics[width=15cm]{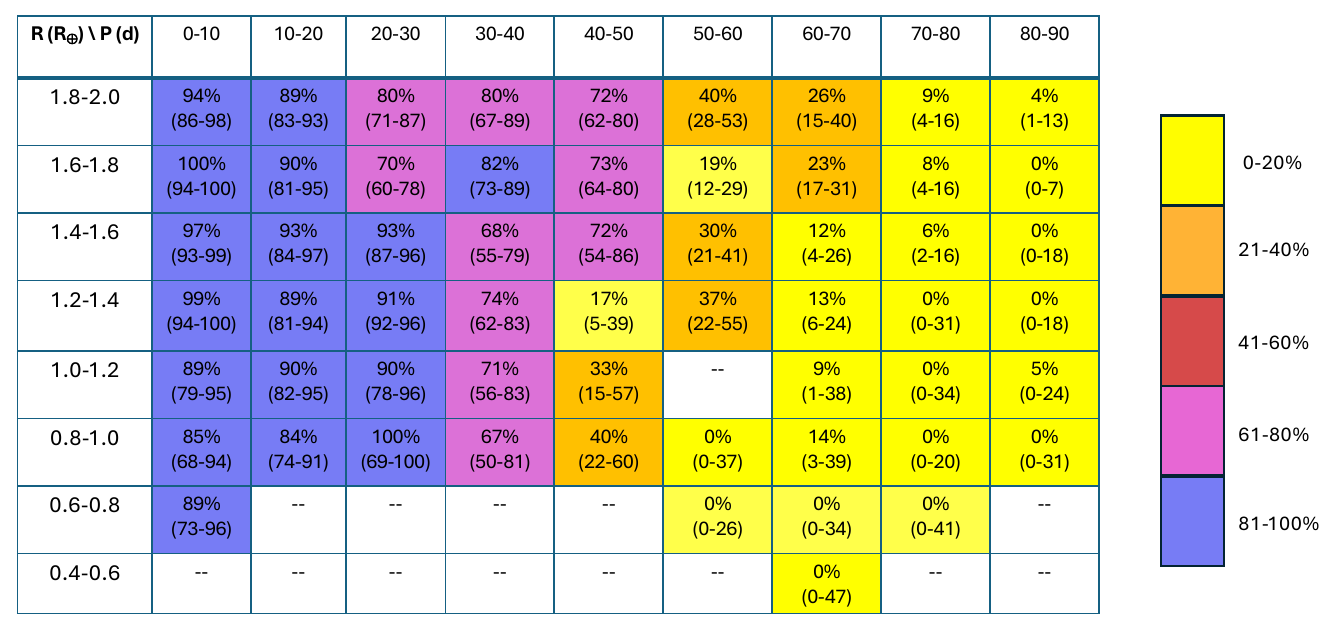}
\caption{Heat map for the reliability of vetted one-quarter PLATO planet using ARPS methodology. Each radius-period cell shows the positive predictive value (equation 2) percentage with 90\% confidence interval.  Fig.~\ref{1Q_recovery_table.fig} gives further details. } \label{1Q_PPV~table.fig}
\end{figure*}

We define ARPS `recovery' of simulated planetary transit signals by cases along the 45~deg line in the right panel of Fig.~\ref{1Q-periods.fig} where the TCF peak period lies within $\pm 0.1$\% of the injected period.  Of the 2712 simulations that pass the vetting thresholds, 1644 (61\%) have recovered periods within $\pm 0.1$\% of the injected periods.  This represents 26\% of the original 1Q simulations. 

The dependence of ARPS recovery in one-quarter simulations is further elucidated in the histograms shown in Fig.~\ref{Four_1Q_hist.fig}.  Recovery rate is largely independent of host star brightness, but depends on stellar radius: recovery falls from 45\% or more for stars with radii below 1.0~R$_\odot$ to less than 15\% above 1.5~R$_\odot$.  The third histogram shows that more than 50\% of injected planets with orbital periods shorter than 20~days are recovered, but the fraction declines to negligible levels for periods exceeding 60~days.  Finally, around 40\% of injected planets with radii larger than 1.5~R$_\oplus$ are recovered, but the fraction declines to 20\% for Earth-size planets and to negligible levels for radii below 0.8~R$_\oplus$.

We provide two heat maps for the one-quarter simulations to measure ARPS methodology sensitivity to simulated PLATO planets.  Figure~\ref{1Q_recovery_table.fig}  measures `recovery rate' as a function of planet radius and orbital period: (number detected) / (number simulated) for each radius-period cell. We do not make the conversion from percentile recovery rates to numbers of planets in the LOPS2 field as this can vary several fold depending on the choice of intrinsic planet occurrence rates \citep{Matuszewski23}. This measure is commonly used to quantify the success of a transit survey \citep[e.g.][]{Howard12, Christiansen20}.   It tells how sensitive PLATO will be for true planets in a chosen radius-period cell.

{As expected, the most easily detectable cases have radii greater than 1.2~R$_\oplus$ with 80\% recovery for periods shorter than 40~d.  For radii above 1.0~R$_\oplus$ about more than 70\% are recovered for periods shorter than 40~d. Recovery drops to levels below 10\% along a line approximately defined by $P~(d) = 50*R^{0.5}~(R_\oplus)$. 

Earth-size planets with radii around 1.0~R$_\oplus$ have particular scientific interest. The ARPS method detects about 70\% of injections of Earth-radius planets for periods shorter than 40~days in one-quarter lightcurves. These represent lower limits to detection fractions as multivariate machine learning classification will outperform our heuristic thresholding of two variables (Appendix C). 

Figure~\ref{1Q_PPV~table.fig} shows a second heat map giving the ratio of (number recovered with correct period) / (number satisfying detection criteria) in each radius-period cell following application of vetting criteria in transit duration and periodogram peak S/N.  In statistical parlance, this is a heat map of
\begin{equation}
PPV = TP / (TP+FP) = 1 - FDR 
\end{equation} 
values where PPV = positive predicted value\footnote{Wikipedia pages \tt{Confusion\_matrix} and \tt{Positive\_and\_negative\_predictive\_values}.}, TP = true positive count, FP = false positive or false alarm count, and FDR = false discovery rate.  This heat map counts (in radius-period cells) successful and failed recoveries of lightcurves in the right panel of Fig.~\ref{1Q-periods.fig}.  It tells the effectiveness of a vetting procedure for reducing false alarms for a chosen planet size and orbital period. Here again the situation deteriorates for smaller planets at longer periods. 

The two heat maps address complementary questions. The TPR heat map tells us how efficiently planets are recovered across the radius-period space.  These values are needed for inferring intrinsic planet occurrence populations from observed transiting samples \citep[e.g.][]{Howard12}.  The PPV heat map tells us how reliable those detections are once the vetting criteria have been applied.   This is most important for ground-based followup: it quantifies the risk that years of large-telescope spectroscopy is devoted to a star where there is no true periodicity present.  Together, these heat maps show the need to characterize both completeness and reliability when assessing PLATO pipeline transiting candidates.  

A third heat map could be constructed from simulations to give detailed astronomical false positive risks, often involving eclipsing binaries.  This complex issue is not treated here but is discussed in \citet{Fressin13, Deleuil18} and \citet{Hadjigeorghiou25} and other studies.

\subsection{ARPS analysis of 2-year simulations} \label{2yr.sec}

Our analysis of the 2-year inner and outer simulations follows closely the analysis of the one-quarter simulations (Sect. \ref{1Q.sec}).  The maximum trial period now extends to 300~days for the inner and 720~days for the outer simulations, rather than 90~days for the single-quarter simulations.  

As with the one-quarter analysis, most of the peak values in TCF periodograms correspond to unrealistic transit durations or periodogram powers consistent with noise (Fig.~\ref{OxPad_cleaning.fig}). Once thresholds on these two variables are applied, most of the ARPS selections with spurious periods are removed (Fig.~\ref{2yr-periods.fig}).   Of the 1441 simulations that pass the two vetting thresholds, 1138 (79\%) have recovered periods within $\pm 0.1$\% of the injected periods.  This represents 28\% of the original 4000 2-year simulations. 

The distributions of ARPS derived planetary parameters for the 2-year simulations (Fig.~\ref{Four_2yr_hist.fig}) are similar to the one-quarter situation shown in Fig.~\ref{Four_1Q_hist.fig} except that, as designed, periods are primarily in the 0--300~day range rather than 0--45~day range. The ARPS planetary radii are often larger and orbital periods are often shorter than the injected properties. Importantly, most of the ARPS transit durations are completely incorrect and not physically possible. Some of the TCF peaks have S/N values that seem insignificant. The left panel of Fig.~\ref{2yr-periods.fig}, like the earlier Fig.~\ref{OxPad_1Q_sim_hist.fig}, shows that most ARPS periods are incompatible with the injected periods. 

But, as in Sect.~\ref{1Q.sec}, most of these problems can be easily identified and eliminated with simple cuts to the ARPS transit durations and TCF S/N ratios.  These are shown in red symbols in Fig.~\ref{OxPad_cleaning.fig}. We can examine the results of cleaning unphysical transit durations and low TCF peak S/N values by comparing the left and right panels of Fig.~\ref{2yr-periods.fig}. For periods shorter than 240 days, 97\% of the TCF spectral peaks correspond correctly to simulated transiting planets. These planets are mostly in the inner simulations with equilibrium surfaces too hot for liquid water.  For periods longer than 360~d; only a minority of simulations have correctly recovered periods.

The histograms in Fig.~\ref{Four_2yr_hist.fig} show a considerable improvement in sensitivity over the single-quarter histograms in Fig.~\ref{Four_1Q_hist.fig}. Over half of simulated transits with periods shorter than 240~d are successfully recovered, decreasing to below 10\% at periods exceeding 400~d. Recovery percentages rise steadily from around 10\% at 0.8~R$_\oplus$ to 55\% at 1.7~R$_\oplus$. The ARPS sensitivity to simulated PLATO transiting planets with 2 year lightcurves is summarized in Fig.~\ref{2yr_recovery_table.fig} showing a heat map of percentage recovery in the orbital period $vs.$ planet radius plane. The upper-left corner of the heat map showing  larger planets orbiting with shorter periods represents the most easily detectable cases with 90--95\% recovery rates. Recovery rates are in the range 40--80\% for sub-Earth-size planets (0.6--1.0~R$_\oplus$) at periods shorter than 60~days but falls off rapidly for longer periods.  Around 360~day period, recovery is around 10--20\% for planets with radii 1.0--1.5~R$_\oplus$. 

For Earth-size planets with radii 1.0~R$_\oplus$, the ARPS methodology applied to 2-year simulated lightcurves recovers approximately 70\% of injections with periods shorter than 2 months dropping to 20\% for periods around 8 months and 10\% for longer periods.  These fractions can be improved by application of multivariate classification procedures (Appendix C). 

The reliability heat map in Fig.\ \ref{2yr_PPV~table.fig} shows very few false alarms for planets with radii as small as 0.4~R$_\oplus$ for periods shorter than 240~day, and still around 50\% reliability out to 720~d. This is good news for selecting targets for ground-based followup of long-period planets. 

\begin{figure*}
\centering
\includegraphics[width=17cm]{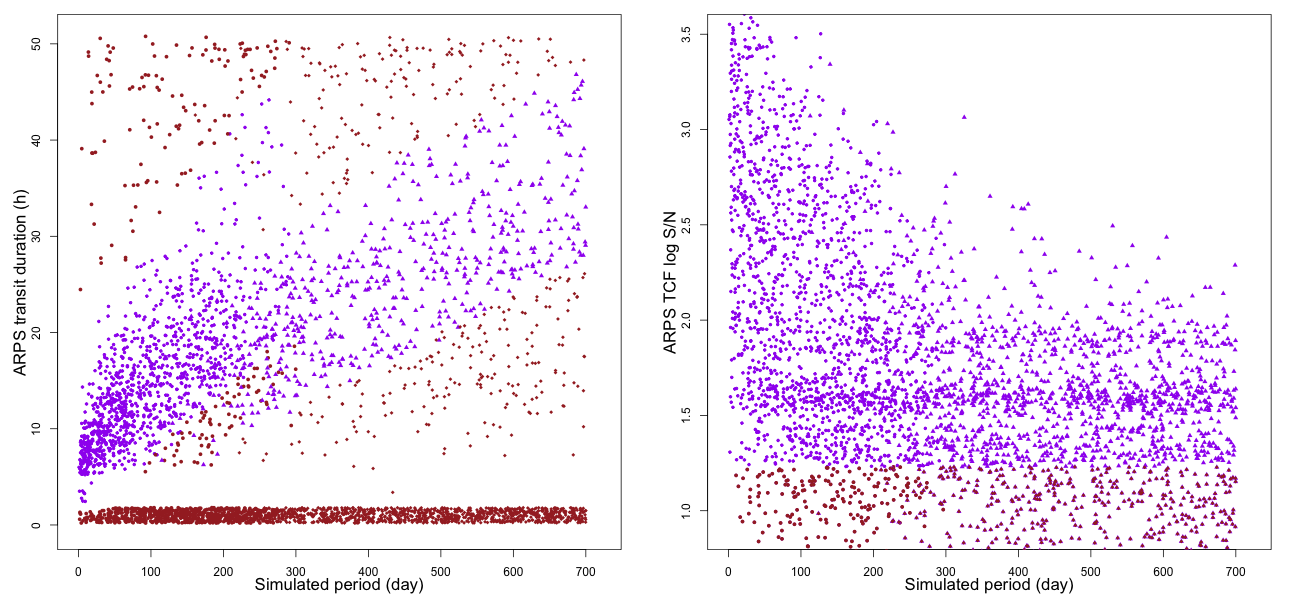}
\caption{Heuristic cleaning of ARPS TCF periodogram best peaks for 2-year simulations.  Fig.~\ref{OxPad_1Q_cleaning.fig} gives further detail regarding the plots.  \label{OxPad_cleaning.fig}}
\end{figure*}

\begin{figure*}
\centering
\includegraphics[width=17cm]{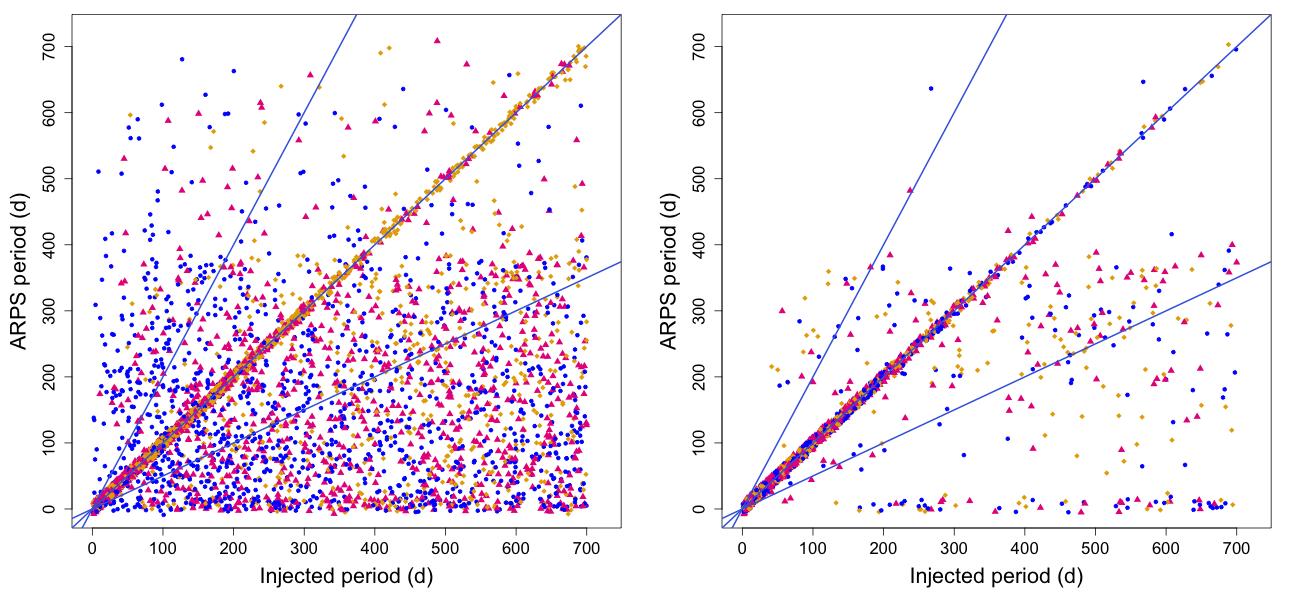}
\caption{Recovered periods $vs.$ simulated periods (in days) for ARPS analysis of 2-year PLATO simulations including both inner and outer simulations.  Left: Full samples from the ARPS processing.  Right: After removal of likely false alarms (red points in Fig.~\ref{OxPad_cleaning.fig}).  Fig.~\ref{1Q-periods.fig} provides detail.  \label{2yr-periods.fig} }
\end{figure*}

\begin{figure*}
\centering
\includegraphics[width=8.8cm]{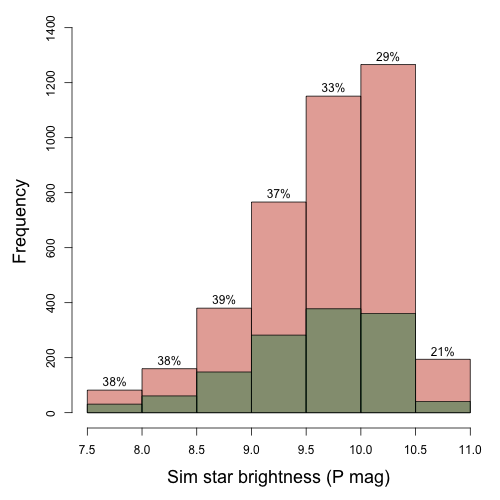}
\includegraphics[width=8.8cm]{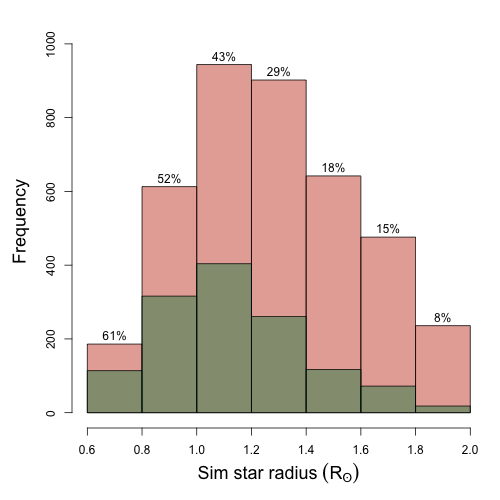} \\
\includegraphics[width=8.8cm]{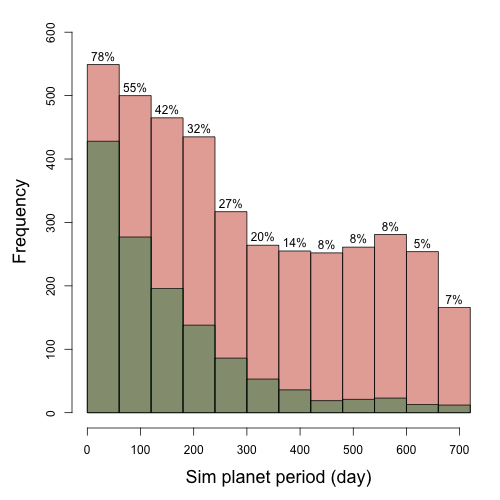}
\includegraphics[width=8.8cm]{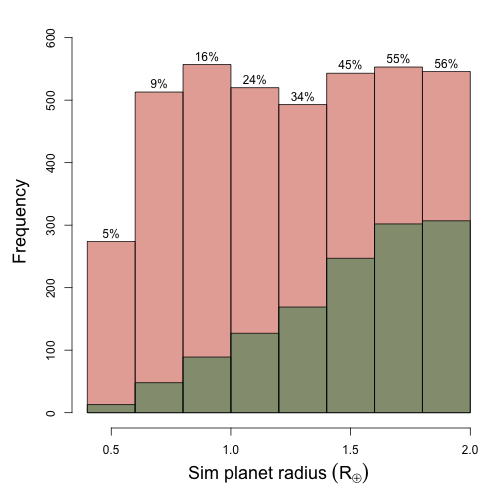}
\caption{ARPS recovery for injected planets for four input parameters of injected planets in the outer simulations observed for the 2-year PLATO pointing. Fig.~\ref{Four_1Q_hist.fig} provides detail. } \label{Four_2yr_hist.fig}
\end{figure*}

\begin{figure*}
\centering
\includegraphics[width=17cm]{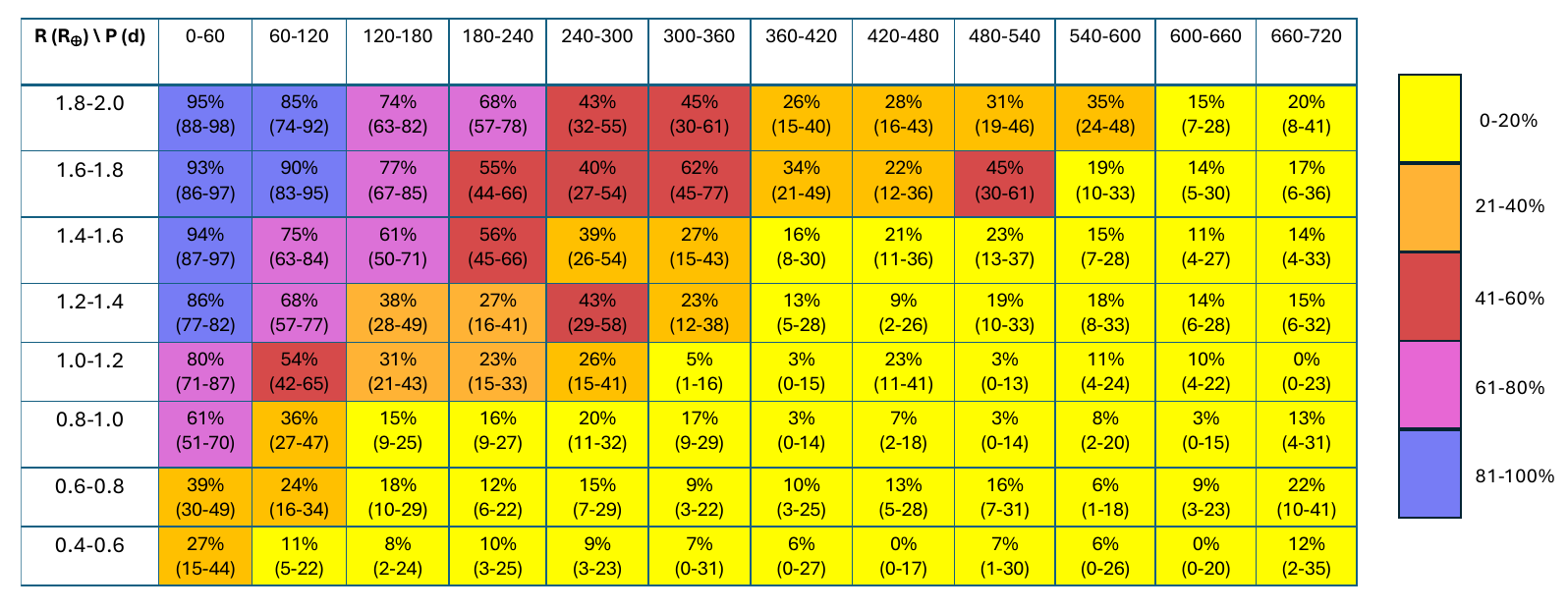}
\caption{Heat map for 2-year PLATO planet recovery using ARPS methodology as a function of orbital period (columns) and planet radius (rows). The sample combines the 2000 inner and 2000 outer simulations.  The values give the percentage of simulated planets recovered based on the heuristic cleaning shown in Fig.~\ref{2yr-periods.fig}. Fig.~\ref{1Q_recovery_table.fig} give details but with a different range of orbital periods.} \label{2yr_recovery_table.fig}
\end{figure*}

\begin{figure*}
\centering
\includegraphics[width=15cm]{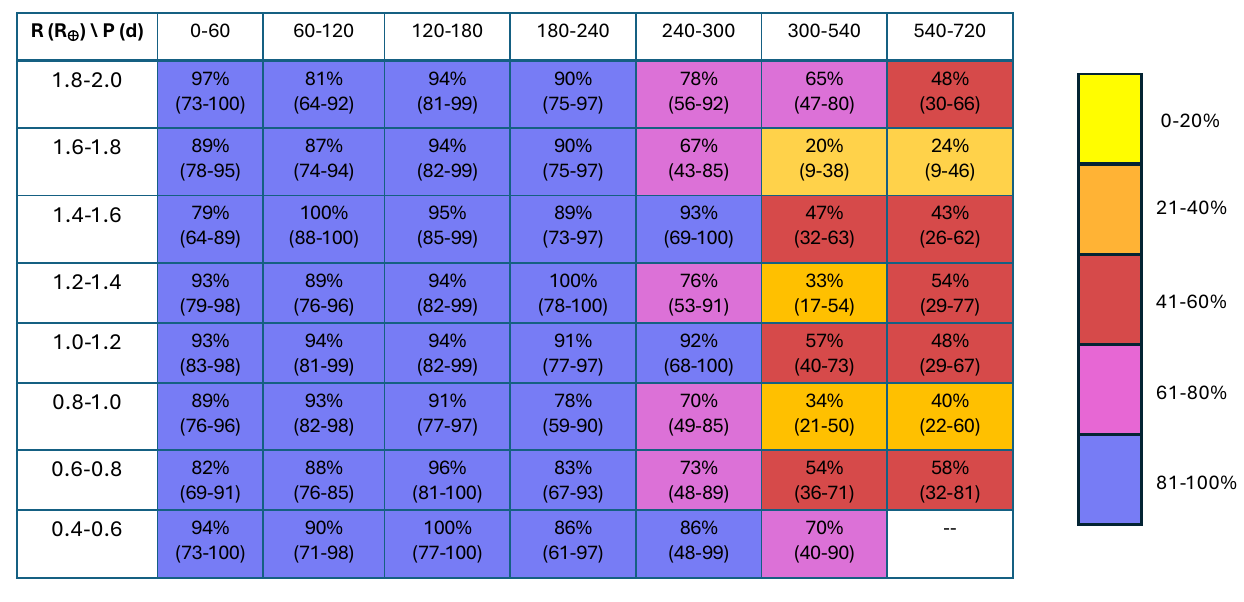}
\caption{Heat map for 2-year PLATO simulations of positive predictive value (equation 2) percentage with 90\% confidence interval using ARPS methodology. Figs.~\ref{1Q_recovery_table.fig} and \ref{1Q_PPV~table.fig} provide further detail. } \label{2yr_PPV~table.fig}
\end{figure*}

\section{Discussion} 
\label{discussion.sec}

\subsection{Summary of findings}

This work investigates the sensitivity of the ARPS methodology for detecting small, Earth-size planets in simulated lightcurves from the planned  PLATO 2-year observation of the LOPS2 field.  The results are promising in many respects. Autoregressive modeling is highly effective in removing stellar variations with different temporal behaviors, as evidenced by the signal-free autocorrelation function in ARFIMA residuals for most simulated lightcurves.  The TCF periodogram, appropriate when differentiation is applied to the original lightcurve for filtering non-planetary signals, is sensitive to Earth-size planets when even when only 0.5\% of photometric observations are in the transits.

After the first 3 months of observations, the ARPS method can detect injected transits from 70\% of planets with radii 1.0~R$_\oplus$ and periods shorter than 40 days. After 2 years of observations, ARPS can  recover approximately 70\% of Earth-like planets with periods shorter than 2 months and 10\% of Earths with periods between 0.8 and 2.0 years.  These main results are shown in heat maps of detection fractions across the exoplanetary period-radius diagram (Figs.~\ref{1Q_recovery_table.fig} and \ref{2yr_recovery_table.fig}). Heat maps relating to False Discovery Rates are also important for predicting contamination by statistical alarms (Figs.~\ref{1Q_PPV~table.fig} and \ref{2yr_PPV~table.fig}). 

Regarding the primary goal of the PLATO mission, the ARPS analysis indicates that a small fraction Earth-like planets in Earth-like orbits can be discovered in a two year survey of the LOPS2 field.  But it is likely these are pessimistic lower limits to the fractions that can realistically be achieved in the 3-month and 2-year PLATO missio.  First, the ARPS methodology has been shown to have weaker performance that other methods because of its detrender: the 'I' differencing operator in ARFIMA modeling is less effective for rapid photometric cadences (Sect. \ref{comparison.sec}-\ref{limitations.sec}). Second, multivariate machine learning classification for construction candidate exoplanet catalogs will always outperform the heuristic vetting procedure used here (Appendix~C).  

The trends shown in  Figs.~\ref{Four_1Q_hist.fig}-\ref{1Q_recovery_table.fig} and \ref{2yr_recovery_table.fig} are reasonable for any planet detection technique.  Transiting signals are more readily recovered for larger planets, shorter periods, brighter stars, and smaller stars.  The detection fractions in the $R-P$ plane shown here can be supplemented by Galactic population estimates to give planetary occurrence rates as functions of stellar and planetary properties \citep{Matuszewski23}.  Preliminary occurrence rates were obtained for ARPS analysis for Year 1 of the {\it TESS} survey by \citet{Melton24c}. 

However, even when the signal is strong, the reliability of discriminating true planets from false alarms (when a periodic signal is incorrectly reported) decreases for planets with periods longer than a third of the lightcurve duration. In Fig.~\ref{2yr-periods.fig} (right panel) for periods in the range 360--720~days, the majority of the cleaned ARPS periods have periods much shorter than the true simulated period.  This problem arises from the deterioration of TCF periodogram when only two or three transits are present as illustrated in Fig.~\ref{app_TCF.fig} and discussed in Sect.~\ref{FA.sec} and Appendix~B.

\subsection{Treatment of false alarms \label{FA.sec}}

Figs.~\ref{1Q-periods.fig} and \ref{2yr-periods.fig} show that controls on transit durations and S/N values greatly reduce the false alarm rates in single-quarter to 2-year simulated lightcurves.  False alarms can be further reduced with multivariate machine learning classification (Appendix C).  However, incorrect periodicities are still present, particularly at periods longer than 1/3--1/2 of the lightcurve duration.  As PLATO mission goals motivate lower classifier thresholds to achieve  higher sensitivity to small planets at longer periods in the recovery heat maps (Figs.~\ref{1Q_recovery_table.fig} and \ref{2yr_recovery_table.fig}), increased false alarms will occur. 

This is an example of the classic trade-off between minimizing Type~I (false positive) and Type~II (false negative) errors in statistical classification. False Discovery Rate (FDR) is a valuable approach to this problem that is widely used in many fields \citep{Benjamini95}.  FDR allows a flexible balance between sensitivity and false alarms where the scientist pre-registers both a desired significance probability (e.g. $p<0.01$) and a desired proportion of false positives such as $FDR = FP / (TP+FP) < 10\%$. FDR is discussed in many reviews \citep[e.g.][]{Benjamini10} as well as volumes like \citet{Efron10} and \citet{Galwey24}.

In this way, PLATO scientists can examine catalogs designed for different scientific purposes.  Catalogs with higher reliability but reduced sensitivity might be used for efficient followup observations while catalogs with higher sensitivity might be used for discovery of more long-period small planets.  In each case, application of the planet detection procedure should be applied to simulations to allow quantitative evaluation of statistical false discoveries (similar to Figs.~\ref{1Q-periods.fig}, \ref{1Q_PPV~table.fig}, \ref{2yr-periods.fig}, and \ref{2yr_PPV~table.fig}) as well as recovery fractions for estimation of planet occurrence rates (similar to Figs.~\ref{1Q_recovery_table.fig} and \ref{2yr_recovery_table.fig}).  These considerations are independent of efforts to reduce false positives from astronomical contaminants like eclipsing binaries  \citep[e.g.][]{Hadjigeorghiou25}. 

Unexpectedly, the recovery heat map (Fig.~\ref{2yr_recovery_table.fig}) shows somewhat different dependency on planetary radii and periods than the false alarm heat map in Fig.~\ref{2yr_PPV~table.fig}.  The recovery heat map shows the expected diagonal pattern where larger planets at shorter periods are more successfully recovered than smaller and longer-period planets.  However, false alarms (the inverse of PPVs) are rare for periods shorter than half the light curve duration, and fall only gradually out to longer durations. The False Alarm rate appears to be independent of planet size.  This is good news for ground-based followup studies, suggesting that Earth-size (and even Mars-size) planet candidates from the PLATO catalogs can be studied without much contamination from statistical false alarms.

\subsection{Comparison with other studies \label{comparison.sec}}

{Our results using the ARPS planet detection methodology can be compared to earlier predictions of PLATO sensitivity to small exoplanets using different detection algorithms (\S 1).  \citet{Krenn24} find that transits of Earth-size planets are much more easily detected if the host star is quiet like the Sun rather than magnetically active.  In contrast, we find that even complex activity signatures can be removed in most cases with effective detrending procedures like ARFIMA(p,d,q) modeling (Fig.~\ref{ARPS_example_good.fig} and Appendix B). 

\citet{Heller2022} use the PSLS set of lightcurve simulations of active stars, a robust (Tukey biweight) sliding window detrender, the Transit Least Squares periodogram that accounts for curvature in the transit ingress and egress shape, and a single vetting signal-to-noise threshold in the periodogram. They obtain recovery rates of approximately 75\% for Earth-size planets (1.0~R$_\oplus$) with two transits orbiting bright stars. This is substantially better recovery than we achieve here with ARPS methodology. Their  recovery rate rapidly falls to 3\% for $R = 0.8$~R$_\oplus$ although it rises to 35\% for shorter periods with three transits in the lightcurve.  This is similar to ARPS performance. 

\citet{Talens26} also obtains greater sensitivity to small planets using other methods. Strong  performance emerges with detrending based on a robust (Huber's M-estimator) spline \citep{Hippke19} followed by the time-domain Cambridge Exoplanet Transit Recovery Algorithm period detection method \citep{Smith25}. Other detection procedures examined by Talens et al. involve various local polynomial regressions filtering and period detection based on limb-darkened transit shapes. 

Our 2-year recovery heat map (Fig.~\ref{2yr_recovery_table.fig}) shows weaker performance than Heller et al. or Talens et al. at long periods.  For Earth-size planets, the ARPS technique needs at least 3 transits for high recovery rates that Heller et al. obtain with $2-3$ transits. Reasons for this failure are outlined in the next subsection.

\subsection{Limitations of ARPS methodology for PLATO lightcurves} \label{limitations.sec}

The overall planetary detection rates obtained here with ARPS methodology are lower than other studies evaluating PLATO sensitivity \citep{Heller2022, Canocchi23, Krenn24, Talens26}. The reason for the weaker performance of ARPS can be attributed to the short cadence time of PLATO.  ARPS was designed for {\it Kepler} and early {\it TESS} data where the photometric cadence was 30~min so that ingress and egress transitions were sudden.  This is essential for the differentiation filtering procedure (the `I' in ARFIMA) to produce a strong transit signal, and the TCF periodogram is used to optimally capture the sharp double-spikes emerging from the ARIMA residuals of a box-shape transit. Thus ARPS was sensitive enough to capture candidate Mars-size planets from {\it Kepler} 4-year data \citep{Caceres19b} and TCF outperformed BLS in {\it TESS} observations of super-Earths due to reduced periodogram noise and trend \citep{Gondhalekar23}.

However, Fig.~\ref{BinWidthTest.fig} shows that the double-spike amplitude in ARFIMA residuals severely weakens as bin widths are shortened towards the native PLATO cadence of 25~s.  ARPS therefore cannot benefit from the useful information in short-cadence lightcurves that reveal the shape of the transit ingress and egress. An astrophysically motivated model of the ingress and egress such as the TLS algorithm \citep{Heller2022} using rapid cadence PLATO data outperform the simplistic box-shape model embedded in ARPS's ARFIMA-TCF procedure.  The planet detection procedures described by \citet{Talens26} are largely independent of cadence bin width because their local regression filtering procedure have characteristic widths of hours, much longer than the difference between 25~s and 30~min cadences that affect ARPS performance. 

\begin{figure}
    \centering
    \includegraphics[width=8cm]{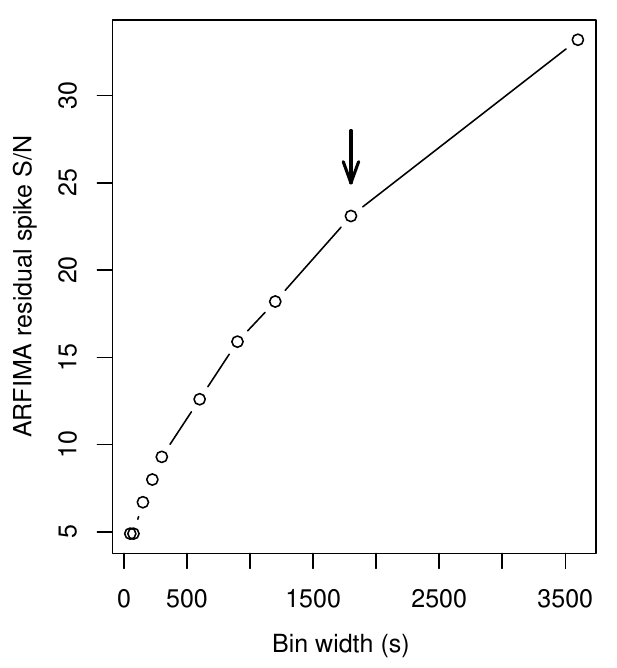}
\caption{The signal-to-noise ratio of the ingress spike in the ARFIMA model residuals as a function of cadence bin widths.  This example is for a simulated super-Earth with period $P \simeq 120$~d and transit duration $Dur \simeq 5$~h. The noise is estimated as $1.48\times$ the median absolute deviation of the detrended lightcurve.  The arrow marks the 30~min bin widths used in the ARPS analysis here whereas the native 25~s cadence of PLATO is the point at the lower left of the diagram. } \label{BinWidthTest.fig}
\end{figure}

\subsection{Implications for PLATO mission sensitivity} \label{implications.sec}

Although the ARPS methodology is not the most sensitive procedure for PLATO planet detection \citep[Sect. \ref{limitations.sec},][]{Talens26}, the analysis here suggests opportunities for improving sensitivity to small planets.

Detrending~~ The differencing operator built into ARIMA and ARFIMA modeling, while effective for {\it Kepler} and early {\it TESS} data, is clearly non-optimal for PLATO with its rapid photometric cadence (Sect. \ref{limitations.sec}). But detrenders based on local regression (such as splines and Gaussian Processes regression) risk a failure to remove rapid, low-amplitude, stationary stochastic variations that may be present.  These can be identified with examining the nonparametric autocorrelation function and applying the Ljung-Box hypothesis tests for short-memory autocorrelation (Figs.~\ref{ARPS_example_good.fig} and \ref{app_ARFIMA.fig}).  If significant stochastic signals are present, then fitting a low-order ARMA model and removing it from the detrended residuals may improve overall sensitivity to the smallest planets.  

Periodic signal detection ~~ The TCF periodogram, like BLS, is also non-optimal for rapid cadence data, as it is designed for sensitivity to box-shaped transits with sudden ingress and egress.  TCF thus misses information embedded in the curved ingress and egress shapes available to techniques such as TLS \citep{Hippke19} and CETRA \citep{Smith25}.  However, the results here suggest that the periodogram noise characteristics at periods longer than 1/3--1/2 times the light curve duration should be carefully examined (Figs.~\ref{ARPS_example_good.fig} and \ref{app_TCF.fig}).  Trends in periodogram median levels should be removed and noise characteristics should be carefully examined, as discussed by \citet{Ofir14} and \citet{Gondhalekar23}, particularly if periodogram signal-to-noise ratios play an important role in planet detection.  

Vetting procedures ~~ False alarms emerging from the periodogram are reduced here using a simplistic and heuristic thresholds applied to two properties of the signal (Figs.~\ref{OxPad_1Q_cleaning.fig} and \ref{OxPad_cleaning.fig}).  But it is far more advantageous to tune a multivariate machine learning classifier that gives optimized results (Appendix C).  The RAVEN system proposed for PLATO is an example of a multivariate machine learning vetting procedure \citep{Hadjigeorghiou25}.  
 
Vetting evaluation ~~ After detrending and periodicity detection, a planet detection pipeline proceeds with a vetting procedure intended to reduce both statistical false alarms and astronomical false positives.  Analysis  of simulated lightcurves with artificially injected planet signals is needed to produce a heat map of pipeline completeness as a function of planet radius and orbital period (Figs.~\ref{1Q_recovery_table.fig} and \ref{2yr_recovery_table.fig}).  Statistically, this is the true positive rate in a classification confusion matrix.    This is essential for converting observed detection rates to intrinsic planet occurrence rates.   But it is also important to examine the heat map of positive predicted value that measures the false alarm rate after detection and vetting has been performed (Figs.~\ref{1Q_PPV~table.fig} and \ref{2yr_PPV~table.fig}).  This is useful for evaluating the false alarm risk of ground-based followup study for PLATO candidate transiting planets.
 
Catalog interpretation ~~ The heat maps in Figs.\ref{1Q_recovery_table.fig}, \ref{1Q_PPV~table.fig},  \ref{2yr_recovery_table.fig},  and \ref{2yr_PPV~table.fig}  show that PLATO catalog of candidate transiting planets emerging regularly from the LOPS2 survey will necessarily have incompleteness that varies with planet size and orbit.  But scientific followup and interpretation may require different true positive rates and positive predictive value levels.  For example, study of hot Earths might tolerate more false alarms to get larger samples than those obtained in catalogs designed for high reliability planet occurrence rate estimation.  The principle of False Discovery Rate classification  \citep[][Appendix C] {Benjamini95} widely used in other fields encourages using different classification thresholds for different scientific purposes.  It is thus beneficial for the PLATO catalog to be accompanied by classification pseudo-probabilities for all PIC stars so scientists can apply other False Discovery Rates as needed for different purposes.

\section{Conclusions}  
\label{sec.conclusions}

We have applied the ARPS -- AutoRegressive Planet Search --  planet detection procedure to simulations of one-quarter and two-year PLATO lightcurves with realistic photometric variations from stellar activity and injected planetary transits.  The ARPS ARFIMA model combines a nonparametric differentiation-based detrender with a parametric linear short-memory autoregressive component and a nonlinear long-memory $1/f^\alpha$ component.  This model is derived from longstanding experience with ARMA-based econometric time series analysis.  The ARFIMA residuals are then examined for box-like periodic variations with the Transit Comb Filter periodogram.  This procedure has been demonstrated to give high sensitivity to small planet detection in Kepler and TESS datasets \citep{Caceres19b, Melton24a}.  

ARPS is not ideal for the native PLATO cadence of 25~s as it is most sensitive when the entire ingress and egress occurs in a single time bin. The ARPS detection rates can thus be viewed as lower limits to methods designed for rapid cadence light curves. 
Despite this limitation, the results give clear indications of PLATO mission capabilities.  For example, in the first quarter (three months) of data, ARPS analysis of P1 and P2 stars in the LOPS2 field predicts detection of a majority of Earth-size planets out to $\sim 40$~days, although the detections at longer periods must combat false alarms in the periodogram.  In two years exposure, at least 25\% of transiting Earths are detected out to $\sim 240$~d with almost no false alarms.  Recovery rates are not highly sensitive to the brightness of the host star but benefit from smaller stellar radii.  

The discussion here also provides guidance for PLATO catalog construction is several respects such as statistical testing of detrender performance, complications in periodogram interpretation, and benefits of machine learning multivariate classification for false alarm and false positive reduction with use of ROC curves and False Discovery Rate corrections.

\begin{acknowledgements}

E.D.F. thanks Nathan Hara (CNRS-LAM), Jon Jenkins (NASA-Ames), Elizabeth Melton (RHIT), and Andrew Pellegrino (Penn State) for thoughtful discussions on methodological issues pertaining to PLATO sensitivies. We are also most appreciative of an anonymous referee whose comments substantially improved this paper. T.Z. acknowledges support from CHEOPS ASI-INAF agreement n. 2019-29-HH.0, NVIDIA Academic Hardware Grant Program for the use of the Titan V GPU card and the Italian MUR Departments of Excellence grant 2023-2027 “Quantum Frontiers”. G.J.T and S.A. acknowledge support from the UK Space Agency as part of its support for the UK component of the PLATO Mission.

\end{acknowledgements}

\bibliographystyle{aa}
\bibliography{ARPS_PLATO}

\appendix

\section{Guidance for software implementation of autoregressive modeling} \label{AppendixA.software}

The software used to perform the ARPS analysis of PLATO simulations is similar to that used by \citet{Caceres19b} and \citet{Melton24a} for {\it Kepler} and {\it TESS} application, respectively.  We only apply two stages of ARPS processing $-$ lightcurve filtering using parametric ARIMA-type modeling and the TCF Filter periodogram $-$ omitting later stages of machine learning classification and vetting for reduction of false alarms and false positives. 

The code is written in the R statistical software environment \citep{RCoreTeam24} but it can be readily wrapped into Python using the {\tt rpy2} interface.\footnote{ {\tt rpy2} code is available at \url{https://rpy2.github.io} with documentation at \url{https://rpy2.github.io/doc/v2.9.x/html}.} The TCF periodogram is written in Fortran and is available from the $Astrophysics~Source~Code~Library$ \citep{Caceres22}.  Following is pseudocode for the ARPS filtering and period search code. Closely related code for a {\it TESS}-oriented ARPS simulation study is provided by \citet{Gondhalekar23}\footnote{\url{https://github.com/Yash-10/Periodogram-
Comparison-Optimize-Planet-Detection.}} and \citet{Gondhalekar23b}. 

The input lightcurve, designated {\tt lc} here, is assumed to have two real-valued columns giving observation time (in evenly spaced bins without gaps) and stellar flux. Missing flux values are permitted (with R entries {\tt NA} representing {\tt Not~Available}) but the ARIMA fitting will deteriorate if missing data are sprinkled throughout the time series.  A few long gaps will not have much effect if the short-memory autoregressive behavior (irrespective of trends) is stationary for the full time series.  

The time-axis bin width should not be much shorter than the expected ingress and egress times for planetary transits.  This is needed to concentrate the ingress and egress signals in a single time bin following the differencing operation. In our experience, satellite cadences should be averaged as needed to give bin widths in the range 20--30 minutes. 

The preparatory code is in the following lines. The {\tt forecast} Comprehensive R Archive Network package in R is the leading autoregressive code for econometric applications.  Described in detail by \citet{Hyndman21}, {\tt forecast} has been downloaded $\sim 7000$ times daily for the past decade.  The {\tt install.packages} function must be run only once in the history of the host computer, but the {\tt library} function for every subsequent R session. The TCF Fortran must be compiled in advance by the user with results placed in directory {\tt tcfdir}. The R function {\tt dyn.load} links the compiled code to the current R session through the function {\tt .Fortran}.  

\begin{verbatim}
install.packages("forecast") 
library(forecast)  
dyn.load(paste0(tcfdir,"tcf.so") 
\end{verbatim}

The first line of the code below performs a single differencing operation (or differentiation) of the stellar flux values that will remove (detrend) most forms of nonstationarity, and simultaneously convert a box shape to a double down-up spike sequence \citep{Caceres19a}.  An initial value of $Not~Available$ is added to maintain the original number of values. The next line performs the ARFIMA({\it p,d,q}) fit; in this case restricted to orders with $p$ and $q$ each limited to 10. Computation is based on maximum likelihood estimation for each order, and the order with the highest value of Akaike Information Criterion is selected as optimal.  \citet{Hyndman21} gives the statistical model and details on the computational procedure.  Overviews of ARMA-type modeling for astronomy are provided by \citet{Feigelson12} and \citet{Feigelson18}. 

The residuals of the ARIMA({\it p,1,d}) fit are captured in the vector {\tt ARresid} in the third line of the following lines.  The autocorrelation function (R function {\tt acf}) of the residuals is then examined and a formal Ljung-Box test for consistency with Gaussian white noise is applied (R function {\tt Box.test}).  

More elaborate autoregressive modeling can be tried if significant residuals are present. In the PLATO simulation analysis here, when the ARIMA residuals exhibit significant structure, we apply a fractional integrated ARFIMA(p,d,q) model where $1.0 < d < 1.5$ is now real-valued.  This is equivalent to adding a long-memory $1/(f^\alpha)$ component to the short-memory ARMA({\it p,q}) component where $\alpha = 2(d -1)$.  Mathematical details are provided by \citet{Palma07}.  

\begin{verbatim} 
diff_lc = c(NA, diff(lc[,2])) 
ARFIMA.fit = arfima(diff_lc,  
     stepwise=T, approximation=T,
     max.p=10, max.q=10, max.d=0) 
ARresid = c(ARFIMA.fit$residuals,NA)
Box.test(ARresid,type="Ljung-Box",lag=10)$p.value
\end{verbatim} \label{ARIMA.R}

The TCF periodogram is then run using the first line of the code below. This is the most computational burdensome step of the ARPS calculation, but typically takes approximately $1$~minute for a 2~year PLATO simulation using a 30~minute time bin width. The Appendix of \citet{Gondhalekar23} discussed the computational time scaling. The periodogram is plotted in the second line, and the peak power period (in cadence number) is obtained in the final lines.  

\begin{verbatim}
tcf_out = tcf(ARresid, p.try=pers) 
plot(tcf_out$inper, tcf_out$outpow, type="l") 
best_power = max(tcf_out$outpow) 
best_period = which(tcf_out$outpow==best_power)
\end{verbatim}

The next code snippet establishes trial periods for which the periodic double-spike matched filter power is calculated for the TCF periodogram following the algorithm proposed by \citet[][Sect. 3.2]{Caceres19a} based on the recommendation of \citet{Ofir14}.  Here the $tcf\_oversample$ can be set to an integer value greater than 1 if more accuracy is desired in the peak periods, at the expense of higher computational cost. The minimum and maximum periods (in cadence number) can be chosen by the scientist.  Here we choose $p_{min} = 10*30$~min to search for periods as short as 5~hours and $p_{max}$ to be the duration of the lightcurve. 

\begin{verbatim} 
tcf_oversample = 2 
N_timebins = length(ARresid) 
p_min = 10 ; p_max = N_timebins 
log_by = log(1+1/N_timebins) / tcf_oversample 
pers = exp(seq(log(p_min),log(p_max),by=log_by)) 
\end{verbatim}

\section{Detailed ARPS performance on PLATO simulations} \label{AppendixB.ARPSperformance}

\begin{figure*}
\centering
\includegraphics[width=15cm]{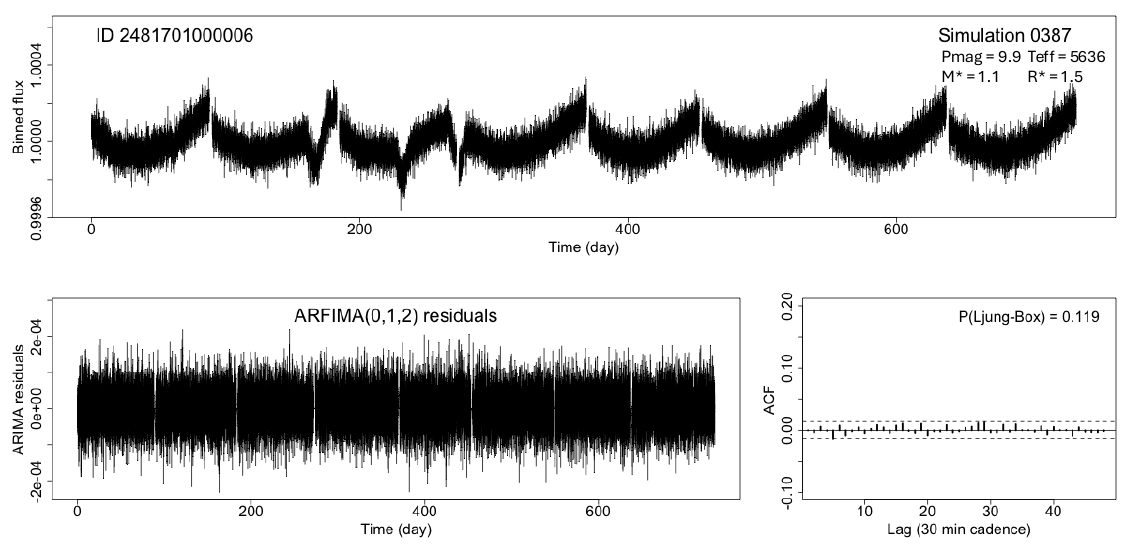}
\includegraphics[width=15cm]{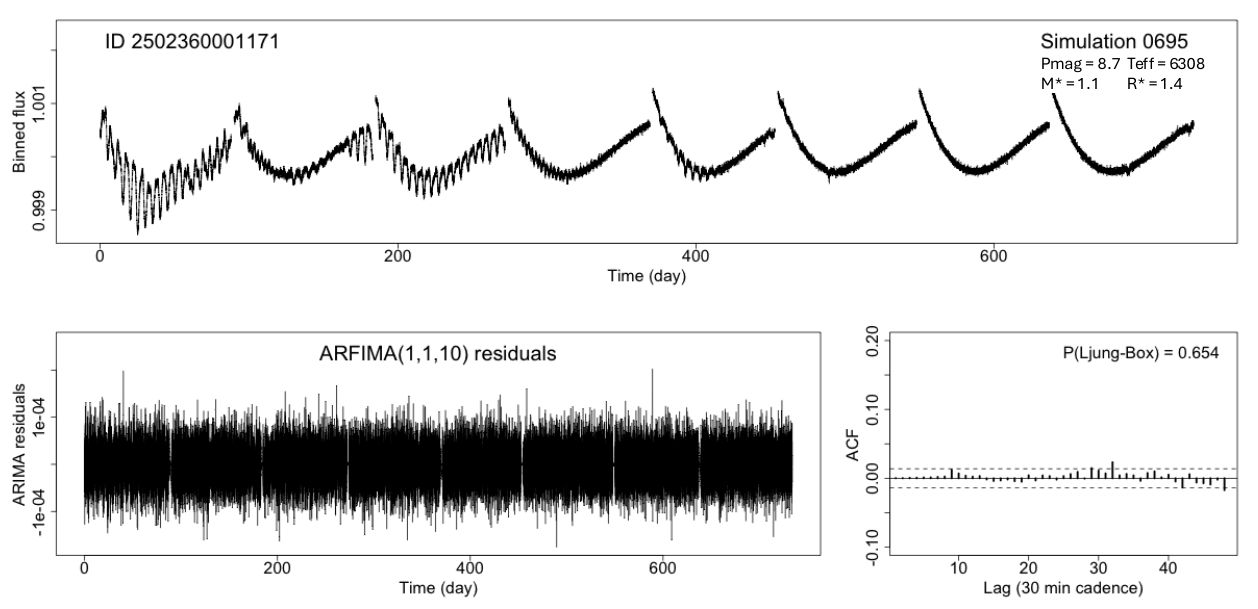}
\includegraphics[width=15cm]{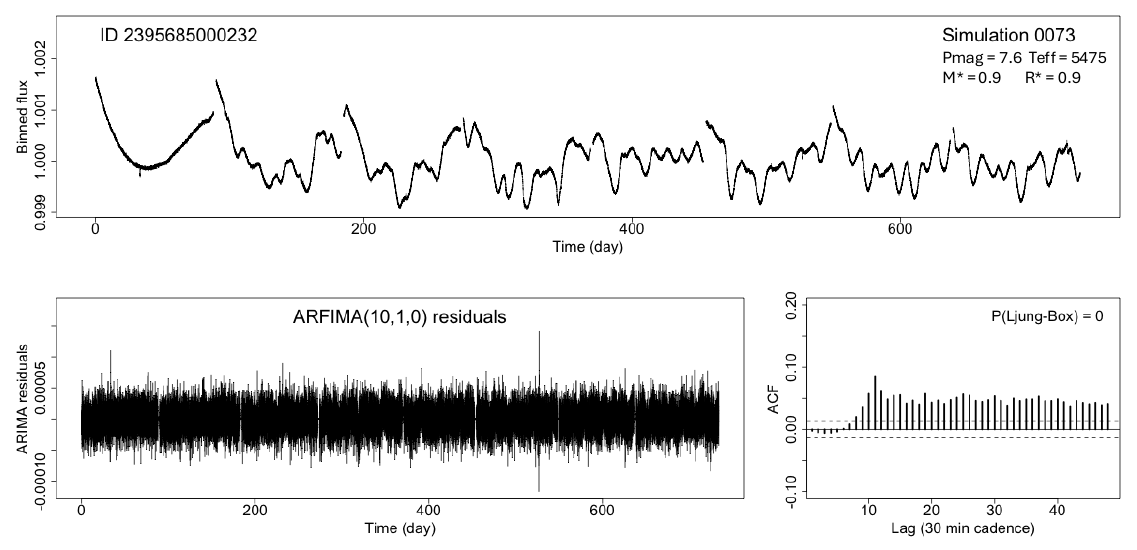}
\caption{Three illustrations of ARFIMA modeling performance on 2-year simulated lightcurves with injected planets.  The text provides details.} \label{app_ARFIMA.fig}
\end{figure*}

\begin{figure*}
\centering
\includegraphics[width=15cm]{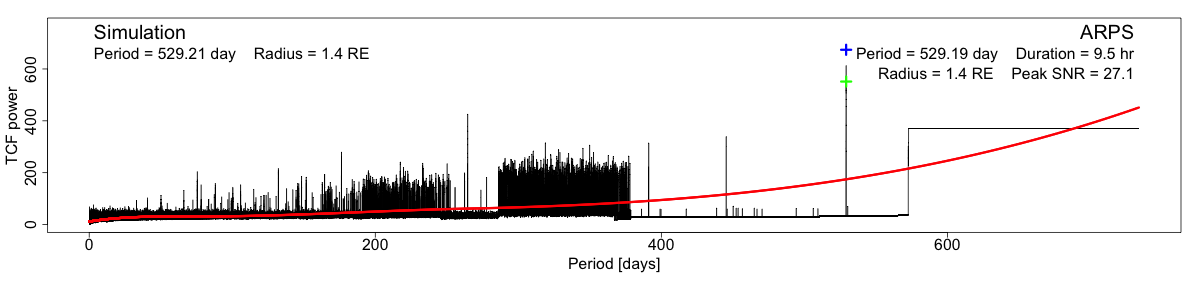}
\includegraphics[width=15cm]{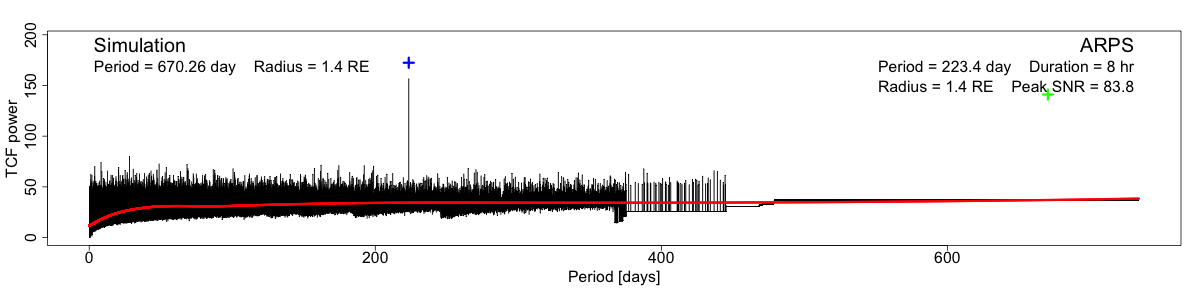}
\includegraphics[width=15cm]{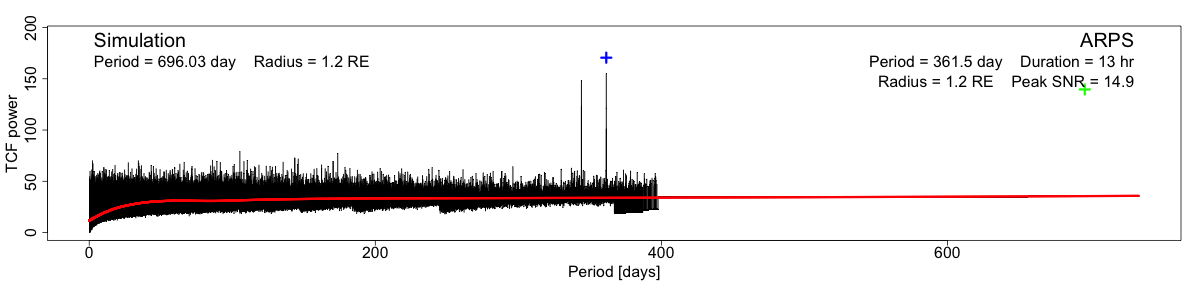}
\caption{Three illustrations of TCF periodogram performance on 2-year outer simulated lightcurves with injected planets. The red curves are LOESS local polynomial smooth fits.  Blue plus marks the peak power period in the TCF periodogram, and green plus marks the simulated transit period.  The text provides details.} \label{app_TCF.fig}
\end{figure*}

As outlined in Sect. \ref{arps.sec}, two stages of the AutoRegressive Planet Search analysis are performed here: ARFIMA modeling of the lightcurve, and TCF periodogram of the ARFIMA residuals.  The best fit ARFIMA({\it p,q}) models chosen with the Akaike Information Criterion (a likelihood ratio test with penalty for model complexity) spanned the range of $0<p<10$ and $0<q<10$ considered in our analysis. Goodness-of-fit is measured with the Ljung-Box test \citep{Ljung78}, a generalization of the Durbin-Watson test that evaluates whether significant autocorrelation is present in the ARFIMA residuals above that expected from Gaussian white noise.  We apply the test out to a lag of 10 cadences (5 hours). 

Figure~\ref{app_ARFIMA.fig} shows two cases where the ARFIMA({\it p,d,q}) model provided an excellent fit to the simulated lightcurves, and one case where the solution was inadequate. The $\simeq$720~day lightcurve in the top two panels has quarter-long variations and a few distinct month-long events between days 170--300.  This behavior was successfully fit with a low-order ARFIMA(0,1,2) model and the residuals are consistent with Gaussian white noise.  The middle panels simulate a rapidly rotating magnetically active star with starspots that emerge and dissipate.  Here a higher-order ARFIMA(1,1,10) model is needed. Although we restricted ARIFA({\it p,d,q}) models to $p\leq 10$ and $q \leq 10$, this model was sufficient to give uncorrelated residuals.  

The bottom panels show a case with more evanescent spots that was partially treated by an ARFIMA(10,1,0). The residual autocorrelation function shows a good fit out to a lag of 10 cadences but poor control at longer lags.  The residual plot shows autocorrelated behavior most clearly around days 320--400.  Despite the limitations of this ARFIMA fit, the injected transit signal for a planet with radius 1.1~R$_\oplus$ and period 493.6~day was the dominant spectral peak in the TCF periodogram. 

When periods shorter than about half of the lightcurve duration are considered, TCF periodograms are densely population with noise. When the periodic transit signal is detected (Figure~\ref{1Q_recovery_table.fig} and \ref{2yr_recovery_table.fig}), a sharp spectral peak appears clearly above the noise. Lightcurves folded at the peak periods show that TCF is remarkably sensitive for 2-year lightcurves, recovering the planetary signal when only $\simeq 0.03$\% of the photometric observations are in the transit.  

But for longer periods approaching the lightcurve duration, the periodogram quality deteriorates with trends (also discussed by Ofir [2014] for Box-Least Squares periodograms) and constant noiseless power levels. A planetary signal is occasionally recovered in this long-period regime but is often missed. 

Figure~\ref{app_TCF.fig} displays TCF periodograms showing one successful recovery of an injected planet and two non-trivial failures.  The top row shows an outer simulation where a simulated planetary signal is recovered in a portion of the periodogram that is not well-populated with noise.  This is a common situation for planets with orbital periods longer than half of the lightcurve duration.  Here tabulated S/Ns become inaccurate and interpretation of signal reliability is difficult.

The middle panel shows a case where an isolated periodogram peak is present with high S/N in a portion of the periodogram well-populated with noise.  But peak period at 223.4 days is the 1/3 alias of the true injected period at 670.3 days. The third panel shows a case where two peaks are present.  The strongest peak is erroneous and the other peak is at half of the true period.    

In these cases, if these were true observations, it would be difficult to know that the identified peaks were erroneous. Perhaps the absence of factor-of-two aliases gives a hint.  Examining alias structure arising from simulated planetary transits injected into the observed lightcurve might be helpful in the case of the outer simulations. Using alias structure in simulations to discriminate misleading spectral peaks from true peaks is recommended by \citet{Dawson10}, \citep{vanderPlas18} and \citep{Getman23} in related circumstances.

\section{The role of machine learning classifiers in transiting exoplanet catalogs}  \label{AppendixC.classifiers}

From a statistical perspective, the reliable selection of a small number of stars with transiting planets from a large collection of photometric lightcurves is a problem in imbalanced classification based on pattern recognition \citep{Fernandez18, Galli26}.  Scientifically successful classification thus requires very small fraction of both false alarms (FAs) where no periodicity is actually present and false positives (FPs) where periodicities have a non-planetary origin. 

In the main body of this study, we consider only detection sensitivity seeking reduction of FAs using heuristic (i.e. not optimized by an algorithm) thresholds on two variables, transit duration and TCF power S/N.  Adding a multivariate machine learning classifier is usually viewed as a tool for optimizing discrimination between transits and FPs.  But it will also substantially improve discrimination between shallow transits and FAs allowing improvements in small-planet sensitivity.  This issue is recently discussed by \citet{Ivashtenko25, Hadjigeorghiou25, Robnik26}.  The heat map recovery rates in Figs.~\ref{1Q_recovery_table.fig} and \ref{2yr_recovery_table.fig} should thus be viewed as lower limits to PLATO’s sensitivity after 1 quarter and 2 years observations, respectively. In the parlance of classification methodology, optimized classifiers will reduce both FAs and false negatives (FNs), thereby improving TP rate or sensitivity (often called ‘completeness’ in astronomical surveys).

Powerful multivariate statistical and algorithmic approaches based on machine learning have been developed in recent decades that can reduce FAs and FNs while simultaneously reducing FP contamination.  Such treatments require: dozens of `features’ characterizing the lightcurve behaviors; large training sets of labeled lightcurves (TPs, FPs, and true negatives or TNs); and well-designed classifier optimization algorithms often involving decision trees, bootstrap aggregation, and gradient descent methods. These methods, developed over several decades \citep[e.g.][]{Schapire90, Breiman01, Quinlan14, Robin11, Chen16}, fall under the rubric of ensemble classifiers. 

Among feature-based machine learning procedures, the Random Forest classification approach \citep{Breiman01} is most commonly used for transiting planet identification.  It has been applied to ground-based, {\it Kepler}, K2, and {\it TESS} mission lightcurves and is a stage of the widely-used NASA-Ames analysis pipeline \citep{McCauliff15, Mislis16, Shallue18, Armstrong18, Schanche19, Caceres19b, Lu20, Montalto20, Armstrong21, Montalto23, Melton24a, Melton24b}.  For exoplanet applications, adaptations to imbalanced classes may be needed  \citep[e.g.,][]{Ishwaran24, smotefamily}. Design of training sets is critical: the DIAmante and DTARPS analyses of {\it TESS} data show that FP contamination can be considerably reduced by introduction of simulated eclipsing binary FPs into the negative training set \citep{Montalto20, Melton24a}. Specialized measures such as comparing in-$vs.$-out of transit  difference images and transit shape classification can also be important \citep{Bryson25}.

An important alternative to Random Forest is XGBoost \citep{Chen16} proposed for PLATO transit discovery by \citet{Hadjigeorghiou25}.  Gradient boosting efficiently focuses on a single best classification while Random Forest will often give a broader range of possible classifications.  XGBoost may thus follow the training set more reliably.  But if one uses training sets based on empirically confirmed transiting planet distributions that are largely missing small planets at long periods, then more flexible approaches like Random Forest might be preferred.

Once a type of ML classifier and a classifier threshold are chosen, trials with different algorithm and parameter choices are compared by confusion matrices, ROC curves, and their associated scalars \citep[such as Accuracy, Precision, Recall, False Discovery Rate, F1 score, Youden's J, and Matthews Correlation Coefficient,][]{Nakas23}.  The classification stage can follow any time series analysis procedures such as those discussed by \citet{Talens26}. 

Fig.~14 of \citet{Caceres19a} illustrates the power of multivariate classification applied to a transiting planet survey. It shows the ROC curve emerging from an optimized Random Forest classifier for {\it Kepler} 4-year lightcurves based on 37 features of the lightcurves, ARPS analysis, and stellar metadata. There is strong improvement (from 40\% to 98\% recovery assuming a FP rate of 1\%) in planet detection sensitivity compared to a univariate threshold on the periodogram peak signal-to-noise ratio.  This indicates that the heat maps derived here based on heuristic thresholds of two variables are pessimistic and that a classifier-enhanced vetting procedure will give higher percentage of TP recovery.

Transiting exoplanet catalogs can furthermore be based on different classifier thresholds.  As the astronomer chooses a lower [higher] threshold for the classifier probability, more [fewer] new planets are discovered with higher [lower] contamination by FPs and FAs.  This tuning of the ratio of FPs to TPs is the principle of False Discovery Rate (FDR) introduced by \citet{Benjamini95} and widely used in many fields.  Rather than applying a conservative Bonferroni correction that minimizes FPs, the FDR framework allows the scientist to choose a ratio that balances TP recovery at the expense of FN, FP and TN contamination.  This choice is based on scientific goals; there is no statistically optimal value. 

Constructing exoplanetary catalogs based on different classifier FDRs (or equivalently, different thresholds along the classifier ROC curve) can serve different scientific purposes.  A conservative classifier, for example, would give a small high-purity catalog with fewer eclipsing binary contaminants but exhibiting reduced ability to discover members of unusual subpopulations such as long-period Earths.  A more liberal classifier might give a larger catalog with higher TP recovery rate and more opportunity for discoveries of members of unusual subpopulations, but with higher FP and FN rates.  

\end{document}